**Title:** Large Eddy Simulation of a Premixed Bunsen flame using a modified Thickened-Flame model at two Reynolds number

**Authors:**

Ashoke De[1], Graduate Student,
Sumanta Acharya [*,1,2], Professor,

[1]Mechanical Engineering Department
Louisiana State University, Baton Rouge, LA 70803

[2]Turbine Innovation and Energy Research Center
Louisiana State University, Baton Rouge, LA 70803

[*]Corresponding Author

Tel.: +1 225 578 5809 Fax: +1 225 578 5924

E-mail address: acharya@me.lsu.edu

**Abstract**

A modified Thickened Flame (TF) model based on Large Eddy Simulation (LES) methodology is used to investigate premixed combustion and the model predictions are evaluated by comparing with the piloted premixed stoichiometric methane-air flame data (Chen et al., 1996, The detailed flame structure of highly stretched turbulent premixed methane-air flames, Combustion and Flame, 107, 233) for Reynolds numbers Re = 24,000 (flame F3) and Re=52,000 (flame F1). The basic idea of Thickened-Flame approach is that the flame front is artificially thickened to resolve on the computational LES grid while keeping the laminar flame speed ($s_L^0$) constant. The artificially thickening of the flame front is obtained by enhancing the molecular diffusion and decreasing the pre-exponential factor of the Arrhenius law. Since the flame front is artificially thickened, the response of the thickened flame to turbulence is affected and taken care of by incorporating an efficiency function (E) in the governing equations. The efficiency function (E) in the modified TF model is proposed based on the direct numerical simulations (DNS) data set of flame-vortex interactions (Colin et al., 2000, A thickened flame model for large eddy simulation of turbulent premixed combustion, Physics of Fluids, 12, 1843). The predicted simulation results are compared with the experimental data and with computations reported using RANS (Reynolds averaged Navier-Stokes) based probability distribution function (PDF) modeling approach (Lindstedt, R. P. and Vaos, E. M., 2006, Transported PDF modeling of high-Reynolds-number premixed turbulent flames, Combustion and Flame, 145, 495) and RANS based G-equation approach (Herrmann, M., 2006, Numerical simulation of turbulent Bunsen flames with a level set flamelet model, Combustion and Flame, 145, 3). It is shown that the results with the modified TF model are generally in good agreement with the data, with the TF predictions consistently comparable to the PDF model predictions and superior to the results with the G-equation approach.

**Nomenclature**

| | |
|---|---|
| A | pre-exponential constant |
| $c_{ms}$ | material surface constant ~ 0.28 |
| $C_s$ | LES model coefficient |
| $C$ | mean reaction progress variable |
| $D_i$ | molecular diffusivity |
| Da | Damköhler number |
| E | efficiency function |
| $E_a$ | activation energy |
| F | thickening factor |
| $h_f^0$ | enthalpy of formation |
| k | thermal conductivity |
| $l_t$ | integral length scale |
| $Pr_t$ | turbulent Prandtl number |
| $Re_t$ | turbulent Reynolds number |
| $s_L^0$ | laminar flame speed |
| $S_{ij}$ | mean strain rate tensor |
| Sc | Schmidt number |
| $Sc_t$ | turbulent Schmidt number |
| $T_a$ | activation temperature |
| $T_b$ | adiabatic flame temperature |
| $T_u$ | temperature of unburnt mixtures |
| $u_i$ | velocity vector |
| $u'$ | rms turbulent velocity |
| $u'_{\Delta_e}$ | sub-grid scale turbulent velocity |
| x$_i$ | Cartesian coordinate vector |
| $Y_i$ | species mass fraction |

**Greek symbols**

| | |
|---|---|
| α | model constant |
| β | model constant |
| $\delta_L^0$ | laminar flame thickness |
| $\delta_L^1$ | thickened flame thickness |
| Δ$_x$ | mesh spacing |
| Δ$_e$ | local filter size |
| Γ | efficiency function |
| μ$_t$ | dynamic turbulent eddy viscosity |
| ν$_t$ | kinematic turbulent eddy viscosity |
| $\overline{\rho}$ | mean density |
| $\tau_c$ | chemical time scale |
| $\tau_i^k$ | sub-grid stress tensor |
| $\tau_t$ | turbulent time scale |
| ω$_i$ | reaction rate |
| Ξ | wrinkling factor |

**Introduction**

Accurate predictions of flame behavior are important for the proper design of combustion systems, such as land based gas turbine combustors, burners for furnaces and boilers, reciprocating engines etc. Since the classical approach of using Reynolds averaged Navier-Stokes (RANS) equations in conjunction with phenomenological combustion models (Poinsot & Veynante, 2001) do not provide accurate predictions, numerical simulations of reacting flows based on the Large Eddy Simulation (LES) technique have been gaining more attention, and are well suited to provide accurate and cost-effective predictions. The main philosophy behind LES of a reacting flow is to explicitly simulate the large scales of the flow and reactions, and to model the small scales. Hence, it is capable of capturing the unsteady phenomenon more accurately than RANS where the entire spectrum of scales is modeled. The unresolved small scales or sub-grid scales must be modeled accurately to include the interaction between the turbulent scales. Since the typical premixed flame thickness is smaller than the computational grid ($\Delta$), the small scale or sub-grid scale modeling must also take care of the interaction between turbulence and the combustion processes. In this paper, we invoked the Thickened-Flame (TF) modeling technique (Colin et al., 2000), where the flame front is artificially thickened to resolve it on computational grid and to define points where the reaction rates are specified using reduced chemical mechanisms. The influence of flame-turbulence interaction in this technique is represented by a parameterized efficiency function. A key advantage of the TF model is that it directly solves the species transport equations and uses the Arrhenius formulation for the evaluation of the reaction rates.

In the present paper, the TF model is modified and its performance explored for high Reynolds number flames. A generic premixed combustion configuration is adopted for which experimental data (Chen et al., 1996) and numerical predictions with other approaches (Hermann, 2006; Lindstedt & Vaos, 2006) are available. The data set considered here contains measurements for premixed flames at Reynolds numbers of 24000 and 52000, denoted as flame F1 and flame F3. Both these flames fall in the thin reaction zone regime, and hence the thickened flame modeling approach is applicable.

In an earlier LES study, an alternative approach using a level-set flamelet model is presented by Duchamp et al. (2000), where they applied the methodology to examine the Bunsen flame at a Reynolds number Re=24000 (flame F3). The reported results show reasonably good agreement for the flow field and turbulent kinetic energy. However, the jet spreading rate and the turbulent kinetic energy in the shear layer are always over-predicted while the mean temperature profiles are shifted radially towards the centerline. No species concentration data are presented.

In the context of RANS based simulations, Prasad and Gore (1999) reported a comparative study of different flame surface density models for Re=24000 (flame F3). They presented the mean flow and temperature profiles, but no species mass fraction and turbulent kinetic energy data were reported. In later studies, Lindstedt and Vaos (2006) reported the use of a PDF transport approach at Re=24, 000 and 52,000, and showed good agreement with data, while Hermann (2006) used a level set flamelet model with different turbulence levels. Their studies show overall good agreement with the measured data except for discrepancies in the turbulent kinetic energy predictions as well as CO predictions.

In turbulent premixed combustion, a popular approach is to rely on the flamelet concept, which essentially assumes the reaction layer thickness to be smaller than the smallest turbulence scales. One of the most popular models based on this concept is the G-equation model. This approach treats the flame surface as an infinitely propagating surface (flamelet), and hence the flame thickness is assumed to be approximately zero (Peters, 2000; Pitsch & Duchamp, 2002). The propagating surface, which is called the flame front, is tracked using a field variable or iso-surface $G_o$. This technique is valid in both the corrugated flamelet regime and the thin reaction zone regime. However, the signed-distance function (which views the scalar G surrounding the front as the signed distance to the front) plays an important role since the G-equation only captures the instantaneous flame front. This dependence on the distance function is an inherent drawback of this method.

Another family of models relies on probability density function (PDF) approach. The probability density function (PDF) is a stochastic method, which directly considers the probability distribution of the relevant stochastic quantities in a turbulent reacting flow. The description of full PDF transport equation approach for turbulent reacting flow has some advantages, e.g., the complex chemistry is taken care of without applying any *ad-hoc* assumptions (like 'flamelet' or 'fast reaction'). Moreover, it can be applied to non-premixed, premixed, and partially premixed flames without having much difficulty. Usually, there are two ways which are mainly used to calculate the PDF: one is *the presumed* PDF *approach*, and other is *the* PDF *transport balance equation approach*. The *presumed* PDF *approach*, which essentially assumes the shape of the probability function P, is simpler to use; however, it has some limitations in the context of applicability. One of the major limitations is not being able to incorporate the higher-moment information systematically into the presumed PDF function which implicitly constrains higher moments (Ihme & Pitsch, 2008). On the other hand, the PDF *transport balance equation approach* solves a transport equation for the PDF function, which is applicable for multi species, mass-weighted probability density function. This method has considerable advantage over any other turbulent combustion model due to its inherent capability of handling any complex reaction mechanism. However, the major drawback of the PDF transport approach is its high dimensionality, which essentially makes the implementation of this approach to different numerical techniques, like FVM (Finite Volume Method) or FEM (Finite Element Method), limited, since their memory requirements increase almost exponentially with dimensionality. Usually, Monte-Carlo algorithms, which reduce the memory requirements, are used (e.g., Pope, 1985). Moreover, a large number of particles need to be present in each grid cell to reduce the statistical error, and this makes it a very time consuming process. So far, the transport equation method has been only applied to relatively simple situations.

In this paper, we use LES based TF modeling approach. The detailed description of this approach is provided in the later section of the paper. In the first part of the paper, the governing equations and the combustion models are presented. Then the concept of thickened flame modeling approach is reviewed, and the modified TF model is presented. Thereafter, details of the numerical methods with the solution procedure are briefly summarized including the effects of the thickening factor and the pilot temperature. Finally, we present the detailed results of the reacting flow simulations for both the flames F1 and F3. These results are compared with the experimental measurements of Chen et al. (1996) and published simulations using RANS based PDF approach and the level-set approach to assess the accuracy of the current model predictions.

**Governing equations and flow modeling using LES**

The filtered governing equations for the conservation of mass, momentum, energy and species transport are given as:

Continuity equation:
$$\frac{\partial}{\partial t}(\bar{\rho}) + \frac{\partial}{\partial x_i}\left(\overline{\rho u_i}\right) = 0 \tag{1}$$

Momentum equation:
$$\frac{\partial}{\partial t}(\overline{\rho u_i}) + \frac{\partial}{\partial x_j}(\overline{\rho u_i u_j}) = -\frac{\partial}{\partial x_i}(\bar{p}) + \frac{\partial}{\partial x_j}\left((\mu+\mu_t)\frac{\partial \bar{u_i}}{\partial x_j}\right) \tag{2}$$

Energy equation:
$$\frac{\partial}{\partial t}(\overline{\rho E}) + \frac{\partial}{\partial x_i}(\overline{\rho u_i}\bar{E}) = -\frac{\partial}{\partial x_j}\left(\bar{u}_j\left(-\bar{p}I + \mu\frac{\partial \bar{u_i}}{\partial x_j}\right)\right) + +\frac{\partial}{\partial x_i}\left(\left(k+\frac{\mu_t C_p}{\Pr_t}\right)\frac{\partial \bar{T}}{\partial x_i}\right)$$
$$+ \frac{\partial}{\partial x_i}\left(\bar{\rho}\sum_{s=1}^{N} h_s \left(D+\frac{\mu_t}{Sc_t}\right)\frac{\partial \overline{Y_s}}{\partial x_i}\right) - \sum_{s=1}^{N} h_{f,s}^0 \overline{\dot{\omega}_s} \tag{3}$$

Species transport equation:

$$\frac{\partial}{\partial t}(\overline{\rho Y_i}) + \frac{\partial}{\partial x_j}(\overline{u_j}\overline{\rho Y_i}) = \frac{\partial}{\partial x_j}\left(\left(D + \frac{\mu_t}{Sc_t}\right)\frac{\partial \overline{Y_i}}{\partial x_j}\right) + \overline{\dot{\omega}_i} \quad (4)$$

where ρ is the density, $u_i$ is the velocity vector, p is the pressure, $E = e + u^2_i/2$ the total energy, where $e = h - p/\rho$ is the internal energy and h is enthalpy, μ is viscosity, k is thermal conductivity, D is molecular diffusivity, $\mu_t$ is turbulent eddy viscosity, $Sc_t$ is turbulent Schmidt number, $Pr_t$ is turbulent Prandtl number, $h_f^0$ is enthalpy of formation, and $\dot{\omega}$ is reaction rate.

To model the turbulent eddy viscosity, LES is used so that the energetic larger-scale motions are resolved, and only the small scale fluctuations are modeled. The sub-grid stress modeling is done using a dynamic Smagorinsky model where the unresolved stresses are related to the resolved velocity field through a gradient approximation:

$$\overline{u_i u_j} - \overline{u_i}\,\overline{u_j} = -2\nu_t \overline{S}_{ij} \quad (5)$$

$$\text{Where} \quad \nu_t = C_s^2(\Delta)^2 |\overline{S}| \quad (6)$$

$$\overline{S}_{ik} = \frac{1}{2}\left(\frac{\partial \overline{u_i}}{\partial x_k} + \frac{\partial \overline{u_k}}{\partial x_i}\right) \quad (7)$$

$$|\overline{S}| = \sqrt{2\overline{S}_{ik}\overline{S}_{ik}} \quad (8)$$

and S is the mean rate of strain. The coefficient $C_s$ is evaluated dynamically (Germano et al., 1991; Smagorinsky, 1963) and locally-averaged.

**Combustion modeling**

Modeling of flame-turbulence interaction in premixed flames requires tracking of the thin flame front on the computational grid. In the present work, we used TF modeling technique, where the flame front is artificially thickened to resolve on computational grid. Corrections are made to ensure that the flame is propagating at the same speed as the un-thickened flame (Poinsot & Veynante, 2001; Colin et al., 2000). The key benefit of this approach, as noted earlier, rests in the ability to computationally resolve the reaction regions and the chemistry in these regions. More details on this approach are described in the following sections.

**Thickened-Flame (TF) Modeling Approach with LES:**

Butler and O'Rourke (1977) were the first to propose the idea of capturing a propagating premixed flame on a coarser grid. The basic idea with this approach is that the flame is artificially thickened to include several computational cells and by adjusting the diffusivity to maintain the same laminar flame speed $s_L^0$. It is well known from the simple theories of laminar premixed flames (Kuo, 2005; Williams, 1985) that the flame speed and flame thickness can be related through the following relationship

$$s_L^0 \propto \sqrt{D\overline{B}}, \delta_L^0 \propto \frac{D}{s_L^0} = \sqrt{\frac{D}{\overline{B}}} \tag{9}$$

where D is the molecular diffusivity and $\overline{B}$ is the mean reaction rate. When the flame thickness is increased by a factor F, the molecular diffusivity and reaction rate are modified accordingly (FD and $\overline{B}$/F) to maintain the same flame speed. The major advantages associated with thickened flame modeling are: (i) the thickened flame front is resolved on LES mesh which is usually larger than typical premixed flame thickness (around 0.1-1 mm), (ii) quenching and ignition events can be simulated, (iii) chemical reaction rates are calculated exactly like in a DNS calculation without any *ad-hoc* sub models, so it can theoretically be extended to incorporate with multi-step chemistry (Colin et al., 2000).

In LES framework, the spatially filtered species transport equation is given in Equation (4), where the terms on the right hand side are the filtered diffusion flux plus the unresolved transport, and the filtered reaction rate respectively. In general, the unresolved term is modeled with a gradient diffusion assumption by which the laminar diffusivity is augmented by the turbulent eddy diffusivity. However, in the TF model, the "thickening" procedure multiplies the diffusivity term by a factor F which has the effect of augmenting the diffusivity. Therefore, the gradient approximation for the unresolved fluxes is not explicitly used in the closed species transport equations. The corresponding filtered species transport equation in the thickened-flame model becomes

$$\frac{\partial \overline{\rho} Y_i}{\partial t} + \frac{\partial}{\partial x_j}(\overline{\rho} Y_i u_j) = \frac{\partial}{\partial x_j}\left(\overline{\rho} F D_i \frac{\partial Y_i}{\partial x_j}\right) + \frac{\overline{\dot{\omega}_i}}{F} \qquad (10)$$

Although the filtered thickened flame approach looks promising, a number of key issues need to be addressed. The thickening of the flame by a factor of F modifies the interaction between turbulence and chemistry, represented by the Damköhler number, Da, which is a ratio of the turbulent ($\tau_t$) and chemical ($\tau_c$) time scales. Da, is decreased by a factor F and becomes Da/F, where

$$Da = \frac{\tau_t}{\tau_c} = \frac{l_t s_L^0}{u' \delta_L^0} \qquad (11)$$

As the *Da* is decreased, the thickened flame becomes less sensitive to turbulent motions. Therefore, the sub-grid scale effects have been incorporated into the thickened flame model, and parameterized using an efficiency function E derived from DNS results (Colin et al., 2000). Using the efficiency function, the final form of species transport equation becomes

$$\frac{\partial \overline{\rho} Y_i}{\partial t} + \frac{\partial}{\partial x_j}(\overline{\rho} Y_i u_j) = \frac{\partial}{\partial x_j}\left(\overline{\rho} E F D_i \frac{\partial Y_i}{\partial x_j}\right) + \frac{E\overline{\dot{\omega}_i}}{F} \qquad (12)$$

where the modified diffusivity ED, before multiplication by F to thicken the flame front, may be decomposed as ED=D(E-1)+D and corresponds to the sum of molecular diffusivity, D, and a turbulent sub-grid scale diffusivity, (E-1)D. In fact, (E-1) D can be regarded as a turbulent diffusivity used to close the unresolved scalar transport term in the filtered equation.

**Proposed modified TF model:**

The central ingredient of the TF model is the sub-grid scale wrinkling function E, which is defined by introducing a dimensionless flame surface wrinkling factor $\Xi$. The efficiency function (E) is defined by the ratio between the wrinkling factor, $\Xi$, of laminar flame ($\delta_L = \delta_L^0$) to thickened flame ($\delta_L = \delta_L^1$) and written as

$$E = \frac{\Xi|_{\delta_L=\delta_L^0}}{\Xi|_{\delta_L=\delta_L^1}} \geq 1 \tag{13}$$

The factor $\Xi$ is the ratio of flame surface to its projection in the direction of propagation, and is defined as:

$$\Xi \approx 1 + \beta \Delta_e \left| \langle \nabla . n \rangle_s \right| \tag{14}$$

where β is a model constant, and $\langle \nabla . n \rangle_s$ is the sub-grid scale surface curvature, which is estimated from the balance equation of sub-grid scale flame surface density. Using the equilibrium assumption (Colin et al., 2000) of the sub-grid scale flame surface density (sub-grid flame stretch, $\langle \kappa \rangle_s = 0$), which balances the production and destruction terms, the following relationship is obtained

$$\langle \nabla.u - nn : \nabla u \rangle_S \overline{\Sigma} = -\langle w \nabla.n \rangle_S \overline{\Sigma} \tag{15}$$

where the destruction is approximated by assuming the flame front displacement speed to be equal to the unstrained laminar flame speed ($s_L^0$) as

$$-\langle w \nabla.n \rangle_S \approx s_L^0 \left| \langle \nabla.n \rangle_S \right| \tag{16}$$

The straining rate is approximated by the following expression

$$\langle \nabla.u - nn : \nabla u \rangle_S \approx \frac{u'_{\Delta_e}}{\Delta_e} \Gamma\left(\frac{\Delta_e}{\delta_L^0}, \frac{u'_{\Delta_e}}{s_L^0}\right) \tag{17}$$

where $\Gamma$ is a fitting function that has to be defined. By combining Eqns. (14)-(17), the flame surface wrinkling factor, Eq. (14), can be written as

$$\Xi = 1 + \beta \frac{u'_{\Delta_e}}{s_L^0} \Gamma\left(\frac{\Delta_e}{\delta_L^0}, \frac{u'_{\Delta_e}}{s_L^0}\right) \tag{18}$$

Hence, to evaluate the factor $\Xi$, the fitting function $\Gamma$ needs to be defined and estimated.

**Proposed fitting function $\Gamma$**

The fitting function, $\Gamma$, can be expressed as a function of the global strain rate ($\langle a_T \rangle_s$), local filter size ($\Delta_e$), local sub-grid scale turbulent velocity ($u'_{\Delta_e}$), and laminar flame speed ($s_L^0$), and is written as

$$\Gamma = \frac{\langle a_T \rangle_s}{u'_{\Delta_e} / \Delta_e} \tag{19}$$

where the effective global strain rate is obtained by integrating the local strain rate over all the scales between the Kolmogorov scale and $\Delta_e$. Hence, to model $\Gamma$, the effective global strain rate needs to be estimated and is obtained by integrating the local strain rate induced due to a pair of vortices.

The strain rate $S_r$ induced by a pair of vortices is expressed as

$$S_r = C_n\left(\frac{r}{\delta_L^1}, \frac{v'}{s_L^0}\right)\left(\frac{v'}{r}\right) \qquad (20)$$

To determine the effective strain rate $S_r$ induced by a pair of vortices of size $r$ and characteristics speed v', the DNS data set (Colin et al., 2000) of classical flame-vortex interaction is used. The DNS data set includes the range $r/\delta^1_L \in [1.2, 30]$ and $v'/s_L^0 \in [0.8, 8]$. The calculation of $S_r$ requires the function, $C_n$, needs to be defined and estimated as follows.

The original $C_n$ function as proposed by Colin et al. (2000) is

$$C_n\left(\frac{r}{\delta_L^1}, \frac{v'}{s_L^0}\right) = \frac{1}{2}\left[1 + erf\left(0.6\ln\left(\frac{r}{\delta_L^0}\right) - \frac{0.6}{\sqrt{v'/s_L^0}}\right)\right] \qquad (21)$$

A modified $C_n$ function with an additional correction tem is invoked in this paper, as suggested by Charlette et al. (2002), and is based on a better parametric fit to the DNS data from Colin et al. (2000). This modified expression contains an additional term, Eq. (22), which ensures that the slow eddies (characteristic speeds less than $s_L^0/2$) do not stretch the flame. Hence, the fitting function, Eq. (19), which is a function of $s_L^0$, and derived from this modified expression, Eq. (22), also does not affect the flame front. Thus the flame surface wrinkling, Eq. (18), which inherently takes care of the interaction between the flame front and turbulence, is also not affected by the slow eddies. Therefore, in this modified TF model, the use of the proposed fitting function, Eq. (22), has two consequences: (a) the eddies smaller than $F\delta_L^0$ do not interact with the flame front, and (b) the eddies with characteristic speeds less than $s_L^0/2$ also do not stretch the flame front. These effects are parameterized in the sub-grid wrinkling factor, Eq. (18). The modified $C_n$ function is written as follows:

$$C_n\left(\frac{r}{\delta_L^1},\frac{v'}{s_L^0}\right)=\frac{1}{2}\left[1+erf\left(0.75\ln\left(\frac{r}{\delta_L^0}\right)-\frac{0.75}{\sqrt{v'/s_L^0}}\right)\right]\times\frac{1}{2}\left[1+erf\left(3\log\left(2\frac{v'}{s_L^0}\right)\right)\right] \quad (22)$$

The flame stretch, Eq. (20), and the expressions, Eqs. (21) and (22), for $C_n$ function are compared in Figs 1 & 2, respectively. The modified $C_n$ function shows better estimates of the flame stretch and to better fit the DNS results.

Charlette et al. (2002) used the similar expression for $C_n$ function, as given in Eq. (22), to model the sub-grid scale wrinkling factor and to come up with their proposed fitting function $\Gamma$. Their analysis was based on the relationship between strain rate and energy spectrum in homogeneous turbulence by making the association between length scale and wave number. In this work, we are following the previous work of Colin et al. (2000), where the effective global strain rate is obtained by integrating the effective strain rate due to a pair of vortices over all the length scales. However, we are using the modified $C_n$ function, Eq. (22), instead of Eq. (21) used in Colin et al. (2000). Considering the Kolmogorov cascade, the velocity v' and length $r$ scales are related as

$$v'=\left(\frac{r}{l_t}\right)^{1/3}u'=\left(\frac{r}{\Delta_e}\right)^{1/3}u'_{\Delta_e} \quad (23)$$

where the velocity u' corresponds to the turbulent integral length scale $l_t$. Hence, the corresponding strain rate becomes

$$\frac{v'}{r} = \left(\frac{l_t}{r}\right)^{2/3} \frac{u'}{l_t} = \left(\frac{\Delta_e}{r}\right)^{2/3} \frac{u'_{\Delta_e}}{\Delta_e} \quad (24)$$

The effective global strain rate integrated over all the scales between the Kolmogorov scale and $\Delta_e$ is given as

$$\langle a_T \rangle_s = \frac{c_{ms}}{\ln 2} \int_{scales} C_n\left(\frac{r}{\delta_L^1}, \frac{v'}{s_L^0}\right) \frac{v'}{r} d\left[\ln\left(\frac{l_t}{r}\right)\right] \quad (25)$$

where $c_{ms}$ is considered as a model constant and the value is given as 0.28 by Yeung et al. (1990). Equation (25) is written as

$$\langle a_T \rangle_s = \frac{c_{ms}}{\ln 2} \frac{u'_{\Delta_e}}{\Delta_e}\left(\frac{\Delta_e}{l_t}\right)^{2/3} \times \int_{\max[\ln(l_t/\Delta_e),0]}^{(3/4)\ln(\text{Re})} C_n\left(\frac{l_t}{\delta_L^1}e^{-p}, \left(\frac{l_t}{\Delta_e}\right)^{1/3}\frac{u'_{\Delta_e}}{s_L^0}e^{-p/3}\right) \times e^{(2/3)p} dp \quad (26)$$

where $\text{Re}_t = l_t u'/\nu \approx 4(l_t/\delta_L^0)(u'/s_L^0)$ is turbulent Reynolds number. The integration is carried out for all the scales below $\Delta_e$. The numerical integration is performed for the modified $C_n$ function, Eq. (22), to calculate $\Gamma$, Eq. (19), and compared with the fitting function $\Gamma_n$, Eq. (27). Figure 3 shows that the $\Gamma_n$ is in close agreement with the numerical integration of $\Gamma$, Eq. (26). The fitting function $\Gamma_n$ is given as

$$\Gamma_n = \left[-0.15\exp\left(-0.15\frac{\Delta_e}{\delta_L^1}\right) - 0.25\exp\left(\frac{u'_{\Delta_e}}{s_L^0}\right) + 0.85\exp\left(-\frac{1.2}{\left(u'_{\Delta_e}/s_L^0\right)^{0.3}}\right)\right]\left(\frac{\Delta_e}{\delta_L^1}\right)^{2/3} \quad (27)$$

Finally, the flame front wrinkling ($s_T/s_L^0$) is expressed as

$$\Xi = \frac{s_T}{s_L^0} = 1 + \beta \Gamma_n \frac{u'_{\Delta_e}}{s_L^0} \tag{28}$$

$$\beta = \frac{2\ln 2}{3c_{ms}\left(\text{Re}_t^{1/2}-1\right)}, c_{ms}=0.28, \text{Re}_t = u'l_t/\nu \tag{29}$$

where $\text{Re}_t$ is the turbulent Reynolds number. The local filter size $\Delta_e$ is related with laminar flame thickness as

$$\Delta_e = \delta_L^1 = F\delta_L^0 \tag{30}$$

For a given $\text{Re}_t$ and the length scale ratio ($\Delta_e/\delta_L^0$), the velocity scale ($u'_{\Delta_e}/s_L^0$) depends only on integral length scale ratio ($l_t/\delta_L^0$). It is observed that for a given $\text{Re}_t$, $s_T/s_L^0$ decreases as the integral length scale increases and reaches to asymptotic values for higher length scale ratios.

The function $\Gamma$ represents the integration of the effective strain rate induced by all scales affected due to artificial thickening. The sub-grid scale turbulent velocity is evaluated as $u'_{\Delta_e} = 2\Delta_x^3 \left|\nabla^2\left(\nabla \times \bar{u}\right)\right|$, and $\Delta_x$ is the grid size. This formulation for sub-grid scale velocity estimation is free from dilatation. Usually, $\Delta_e$ differs from $\Delta_x$, and it has been suggested that values for $\Delta_e$ be at least $10\Delta_x$ (Colin et al., 2000).

The parameterization discussed above in both the original TF model (Colin et al., 2000) and the present modified TF model are derived based on the DNS data sets of 2D classical flame-vortex interaction results (Colin et al., 2000). Using a broader range of data sets that incorporate swirl, separation and other real effects for defining the TF model parameters would be appropriate, but such data sets that provide the needed turbulent-chemistry correlations are not available. However, other groups have used the original TF model to simulate large scale combustor problems with swirl and coaxial shear (Roux et al., 2005; Selle et al., 2004) and have achieved acceptably good results providing a basis of using the defined TF parameters for a broader class of problems.

**Chemistry model**

As all the species are explicitly resolved on the computational grid, the Thickened Flame model is best suited to resolve major species. Intermediate radicals with very short time scales (compared to flow motions) are computationally difficult to resolve since such resolution may induce a prohibitive thickening to major species. To resolve intermediate radicals may need a substantially larger computational grid. Therefore from the perspective of balancing accuracy and computational economy, a two-step reaction chemistry is explored in the present work.

A two step chemistry, which includes six species ($CH_4$, $O_2$, $H_2O$, $CO_2$, $CO$ and $N_2$), is given by the following equation set.

$$CH_4 + 1.5 O_2 \rightarrow CO + 2 H_2O$$
$$CO + 0.5 O_2 \leftrightarrow CO_2$$

The corresponding reaction rate expressions are given by:

$$q_1 = A_1 exp(-E^1_a/RT)[CH_4]^{a1}[O_2]^{b1} \qquad (31)$$

$$q_2(f) = A_2 exp(-E^2_a/RT)[CO][O_2]^{b2} \qquad (32)$$

$$q_2(b) = A_2 exp(-E^2_a/RT)[CO_2] \qquad (33)$$

where the activation energy $E^1_a$ =34500 cal/mol, $E^2_a$ =12000 cal/mol, $a_1$=0.9, $b_1$=1.1, $b_2$=0.5, and $A_1$ and $A_2$ are 2.e+15 and 1.e+9, respectively, as given by Selle et al. (2004). Properties including density of mixtures are calculated using CHEMKIN-II (Kee et al., 1989) and TRANFIT (Kee et al., 1986) depending on the local temperature and the composition of the mixtures at 1 atm. In the literature by Selle et al. (2004), they have compared this two-step scheme with GRI mechanism and showed that this two-step mechanism yields fairly good comparison with respect to measured temperatures, burnt gas temperatures, major species mass fractions and flame speeds for a range of fuel-air ratios.

**Solution Procedure**

In the present study, an in-house parallel multi-block compressible flow code for an arbitrary number of reacting species, in generalized curvilinear coordinates is used. Chemical mechanisms and thermodynamic property information of individual species are input in standard Chemkin format. Species equations along with momentum and energy equation are solved implicitly in a fully coupled fashion using a low Mach number preconditioning technique, which is used to effectively rescale the acoustics scale to match that of convective scales (Weiss & Smith, 1995). An Euler differencing for the pseudo time derivative and second order backward 3-point differencing for physical time derivatives are used. A second order low diffusion flux-splitting algorithm is used for convective terms (Edwards, 1997). However, the viscous terms are discretized using second order central differences. An incomplete Lower-Upper (ILU) matrix decomposition solver is used. Domain decomposition and load balancing are accomplished using a family of programs for partitioning unstructured graphs and hypergraphs and computing fill-reducing orderings of sparse matrices, METIS (Karypis & Kumar, 1999). The message communication in distributed computing environments is achieved using Message Passing Interface, MPI. The multi-block structured curvilinear grids presented in this paper are generated using commercial grid generation software GridPro$^{TM}$.

**Flow configuration**

The configuration of interest in the present work is the Bunsen burner geometry investigated by Chen et al. (1996), is shown in Figure 4. The flame is a stoichiometric premixed methane-air flame, stabilized by an outer pilot. The incoming streams of both the main and pilot jets have the same composition. The main jet nozzle diameter (D) is 12 mm. The pilot stream is supplied through a perforated plate (1175 holes of 1 mm in diameter) around the main jet, with an outer diameter of 5.67D. The Reynolds numbers used in the present work are Re=24,000 (flame F3) and Re=52,000 (flame F1). Based on the estimated characteristic length and time scale given in Chen et al. (1996), flame F1 & F3 correspond to the thin reaction zone regime.

The computational domain extends 20D downstream of the fuel-air nozzle exit, 4D upstream of the nozzle exit and 4D in the radial direction. The LES grids studied consists of 300x94x64 grid points downstream of the nozzle exit plus 50x21x64 grid points upstream, and corresponds to approximately 1.88M grid points. At the inflow boundary, the instantaneous velocities are computed using a random flow generation technique (Smirnov et al., 2001). Convective boundary conditions (Akselvoll & Moin, 1996) are prescribed at the outflow boundary, and stress-free conditions are applied on the lateral boundary in order to allow the entrainment of fluid into domain. The time step used for the computation is dt=1.0e-3, and the heated pilot temperature is chosen to be 2005K.

To examine grid independence, we have investigated two different grids for cold flow analysis: one is the present grid with 1.88M grid points (coarse) and other one with 5.91M grid points (fine). Both the fine and coarse grid predictions properly capture the spreading rate of the jet and corresponding evolution of the turbulent kinetic energy, and the results are in good agreement with each other and measurements, as shown in Fig. 5. Hence, the coarser grid with 1.88M gird points is chosen here for the more expensive reacting flow calculations. The grid resolution in the computational domain with 1.88M grid points is given as: (a) $\Delta x$=0.08 mm is maintained throughout the whole domain starting from jet inlet to outlet, (b) along the central jet

Δr=0.047 mm and Δθ=0.05 mm is maintained, (c) in the shear layer Δr=0.047 mm and Δθ=0.17 is maintained, and (d) finally at the lateral boundary Δr=0.06 mm and Δθ=0.47 mm is maintained. More detailed results of the grid independence study are presented in (De & Acharya, 2008).

**Results and Discussion**

In this section we present the predicted simulations from the LES-modified TF model and compare with the measured data by Chen et al. (1996) as well as with the computations reported using RANS based probability distribution function (PDF) modeling approach by Lindstedt et al. (2006) and RANS based G-equation approach by Hermann (2006). In evaluating the models, the perspective of computational economy must be kept in mind. A multi step calculation requires the calculation of transport equations for multiple species, and with a PDF modeling approach; the reaction rate expressions require the use of a look-up table with substantial computational input/output (I/O) requirements. In the Thickened Flame modeling approach, the reaction rate expressions are computed using Arrhenius law, and hence computational effort including I/O time is minimized. In terms of CPU costs using TF model, the efficiency for solving 10 PDEs in a fully coupled implicit fashions (reacting flow computations) with 1.88M grid points on a system (15.322 TFlops Peak Performance with 2 Dual-Core 2.66 GHz Intel Xeon 64bit processors and 4 GB RAM per node) is given as: (a) running on 1 processor takes 1703sec/iteration (773.26 microsec/pt/iteration), (b) running on 8 processors takes 48.91sec/iteration (22.21 microsec/pt/iteration), and (c) running on 64 processors takes 8.022sec/iteration (3.64 microsec/pt/iteration).

In presenting the results, we will discuss predictions of the mean axial velocity, turbulent kinetic energy, mean temperature, and species mass fraction. In the first part of this section, we will first examine the sensitivity of the predictions to the choice of two global parameters: the thickening factor F and the pilot temperature. This will be followed by the results from the detailed simulations for low and high Reynolds numbers.

**Comparison of modified TF model and original TF model**

The original TF formulation is based on the $C_n$ function, Eq. (21), as given by Colin et al. (2000). Figures 6-7 show the comparison of the predicted results using two different TF model formulations for flame F1 & F3. Both the TF models produce very comparable predictions, but with some modest improvements with the TF Model proposed in this paper. For flame F3, it is clearly observed that original TF model under-predicts the mean axial velocity in the shear layer region and show greater radial spreading at the downstream locations. For flame F1, the under-prediction by the original TF model is also visible, but greater in magnitude at the location X/D=8.5, as shown in Fig. 7.

**Effects of Thickening Factor (Re=24000)**

Figure 8 shows the predicted results using two different thickening factors F (=10 and 20) for flame F3 (Re=24000). The thickening factor is a parameter in the TF model that modifies the mean reaction rate and the diffusion terms in the governing equations. However, increasing the thickening factor reduces the Damköhler number, and thus the flow field becomes less sensitive to the turbulence motion. It is observed that the two different thickening factors do not show any major differences in predictions except for the turbulent kinetic energy predictions (Fig 8) at the location X/D=6.5 where it is slightly over-predicted with the higher F. The kinetic energy predictions show good agreement with data at F=10. As expected, as the thickening factor is increased, the reaction rate is decreased and in turn reduces the mean temperatures. As the temperature levels go down, it reduces the local viscosity and increases the kinetic energy predictions. While no specific choices for thickening factor have been proposed in the literature, it has been suggested that the effect of thickening factor becomes significant for the values larger than F=10 (Colin et al., 2000). Hence, based on the comparisons made here, the thickening factor F=10 would be an appropriate choice, and therefore, further calculations reported are based on the thickening factor F=10.

**Effects of pilot temperature (Re=52000)**

The effects of different pilot temperatures have been studied and reported in the Figure 9 for flame F1 (Re=52000). The mean progress variable is defined as $C= (T-T_u)/(T_b-T_u)$, where the un-burnt temperature $T_u$= 298 K and the burnt temperature $T_b$=2248 K. The figure includes the comparison between two different temperature boundary conditions for the heated pilot: one is 2005 K and the other one is 1785 K. No specific pilot temperature data is prescribed in the measurements which can be used as boundary condition for computations. Therefore, the pilot composition was computed based on chemical equilibrium where the enthalpies are taken for 10% and 20% heat loses to burner, corresponding to pilot temperatures are 2005 and 1785K (Lindstedt & Vaos, 2006). It is clearly observed that the different pilot temperature boundary conditions produce very comparable predictions except for the predicted temperature in the flame region which is slightly lower in the case of lower pilot temperature, i.e. 1785 K. Hence, from this point we will present predictions using the pilot temperature boundary condition as 2005 K.

**Modified TF model Results-Detailed Distributions**

Based on the observations so far, additional results will be presented with the modified TF model using the thickening factor F=10 and pilot temperature boundary condition as 2005 K. In this section, attention will be focused on detailed comparisons of the modified TF model, the RANS based PDF and the G-equation model and the experimental data for the mean velocity, temperature, kinetic energy and all major species. This comparison will be presented at several X/D locations (2.5, 4.5, 6.5, and 8.5).

It should be noted that the present LES simulations are compared with published simulations from other RANS based model predictions with the PDF approach (Lindstedt and Vaos, 2006) and G-equation approach (Hermann, 2006) for the same flame data at the two Reynolds numbers (24,000 and 52,000). At the lower Re, LES predictions with the G-equation approach (Duchamp and Pitsch, 2000) are also available, and we have also included this in the comparison. However it should be noted that the various simulations presented use different numerics and chemistry compared to the present simulations. Therefore, a one-to-one comparison is not possible between all these models, and the performance of the various models should be viewed from this perspective.

*<u>Mean axial velocity</u>*

Figures 10 & 11 show the mean axial velocity profiles at different axial locations for Re=24000 (flame F3) and Re=52000 (flame F1), respectively. The overall velocity predictions using modified Thickened Flame model are in reasonably good agreement with experimental data as well as with the PDF model predictions. At the lower Re, the TF model predictions show a slightly greater spreading in the mixing layer. At the higher Re, both the TF and PDF model predictions are in excellent agreement with the measurements. The RANS based G-eqn model

shows over-predictions particularly for the lower Reynolds number case. Similar trends are also observed with the LES based G-eqn simulations as reported by Duchamp et al. (2000).

When compared with cold flow predictions (Fig. 12), the radial profiles of the mean axial velocity are broadened in the reacting case, due the effect of the flame front, pushing the shear layer outward in the radial direction. Furthermore, it is observed that the peak center line velocity remains almost constant in the axial direction, and exhibits a longer potential core compared to the cold flow case. These effects are reasonably well reproduced by the present simulations.

*Turbulent kinetic energy*

Turbulent kinetic energy predictions for the reacting flow using the modified TF model are shown in Figures 13 and 14. Figure 12 shows the comparison between the reacting and non-reacting kinetic energy profiles at X/D=4.5 and 6.5. In general, the agreement between the measured and predicted kinetic energy profiles is excellent. It is observed that the measured and predicted turbulent kinetic energy along the centerline increase at the downstream for the non-reacting case while it remains almost constant for the reacting case. Also, the peak value of kinetic energy decreases in the axial direction for the non-reacting case, whereas it increases for the reacting case. Furthermore, for the reacting case, the kinetic energy peak moves away from the centerline further downstream (as the jet entrains air), while it moves towards the centerline for the non-reacting case (Fig. 12).

The kinetic energy simulations for the reacting flow clearly show differences compared to the measured data. For flame F3 (Re=24000), the general trends are well predicted with the TF model till about X/D=6.5. The PDF model consistently shows over-prediction of the turbulence levels at all radial locations and particularly upto X/D=6.5. The RANS-based G-equation model shows reasonable agreement till X/D=4.5 and then shows under-prediction at the downstream locations (X/D=6.5 and X/D=8.5) for r/D>0.5. However, it should be noted that the LES coupled G-eqn simulation (Duchamp & Pitsch, 2000) shows over-predictions in the kinetic energy at all the locations excluding X/D=2.5 where it shows better agreement. The differences in the kinetic energy predictions between the LES results in (Duchamp & Pitsch, 2000) and the RANS results in (Hermann, 2006) are presumably linked to the turbulence modeling issue.

At the higher Reynolds number, for flame F1 (Re=52000), the present simulations are in good agreement with data and capture the general trend quite well, even at the downstream location of X/D=8.5 where, at the lower Re, the TF model under-predicted the measurements. At the location X/D=2.5, close to nozzle exit, the modified TF model under-predicts the turbulence levels in the inner edge of shear layer; however, the model predictions appear to improve at the downstream locations (X/D=4.5, 6.5 and 8.5) where the general trends and peak values are well predicted. In comparing the different model predictions, the TF and PDF model predictions show good agreement with the data while the RANS based G-eqn model under-predicts the peak values by 35-50% at the downstream locations of X/D=6.5 & 8.5. These discrepancies are explained using the mean temperature predictions and discussed in the following sub-section. As observed clearly, in assessing different model predictions, both the modified TF model and PDF model predictions appear to provide reasonably good agreement with the data.

Comparing the TF model predictions at different Reynolds numbers, it is evident that the modified TF model appears to improve the results at the higher Reynolds number. This means the efficiency function parameter E in the governing Eqn. (12) is better suited to properly estimate the flame-turbulence interaction at higher turbulence intensity. This issue of accurately parametrizing the efficiency function needs more investigation.

*Mean temperature*

Mean temperature profiles obtained for different Reynolds numbers are presented in Figs. 15 and 16. The mean progress variable is defined as $C= (T-T_u)/(T_b-T_u)$, where $T_u$= 298K and $T_b$=2248 K. It is immediately evident from the measured temperature distributions at the two Reynolds numbers that the maximum temperature is reduced as the jet velocity (Reynolds number) is increased. This fact can be explained by the fact that as the jet velocity increases, more cold ambient air is entrained on the bunt side of the flame, thus reducing the maximum temperature. This behavior is seen consistently in both the data and the model predictions.

For flame F3 (Re=24000, Fig. 15), as observed immediately downstream of the nozzle exit (X/D=2.5), the mean temperature profile shows an over-prediction. This is likely due to the effect of the chosen 2-step chemistry model, which is based on a reduced complex chemistry and very much likely to over-predict temperature in the fuel-rich regions. However, at the downstream locations (X/D=4.5, 6.5 and 8.5) the mean temperature predictions are in better agreement with the data. In particular, both the modified TF model and the PDF model predictions show good agreement, while both the RANS and LES based G-equation simulations consistently show an over prediction. At all the X/D locations, the G-eqn temperature predictions exhibit a radial shift of the shear layer towards the centerline. One deficiency in the TF model predictions at the lower Re is that the spreading of the temperature shear layer is under-estimated, and this is most likely due to the artificial thickening of the flame front and the diffusivity in the shear layer. Also, it is noticed that the downstream centerline mean temperature in all the simulations excluding LES based G-eqn simulation remains almost close to the un-burnt temperature, whereas it increases in the experiment. For the TF model, this may be linked to the under-prediction of the kinetic energy (and the turbulent diffusivity) near the centerline observed in Fig. 13.

In Figure 16 for flame F1 (Re=52000), there is a greater over prediction of the maximum temperature by the modified TF model at X/D=2.5; further downstream, for higher values of X/D the mean temperature profile is captured very well. Comparing all the model predictions, the PDF model consistently shows over-predictions at all the axial locations, whereas the RANS based G-eqn model shows overall better agreement with the data excluding the locations X/D=2.5 & 4.5 where it shows under-prediction of the maximum temperature. The over-prediction by the modified TF model in the near-field, as noted earlier, is possibly due to the chosen chemistry scheme, which is likely to produce higher flame temperature in the fuel rich regions (close to the nozzle exit). Surprisingly, the PDF model predictions consistently show high temperature distributions in spite of using more detailed chemistry calculations in their model and yielding reasonably good turbulent kinetic energy predictions though (Fig. 14). On the other hand, the RANS based G-eqn shows overall better agreement in predictions of temperature profiles, but as seen in Fig. 14, the turbulent kinetic energy predictions exhibited over-predictions in the near field and lower peak kinetic energies in the far field.

Comparing different Reynolds numbers, it is clearly observed that the modified TF model appears to improve the results for higher Reynolds number, capturing both the trends as well the magnitude of the maximum temperature. As the Reynolds number increases, the maximum temperature also decreases due to the greater entrainment of cold ambient air driven by the higher jet velocity. Compared to the lower Reynolds number (Re=24000), where the thermal shear layer was thinner than in the data, the temperature profile at the downstream location (X/D=8.5) is properly predicted for the higher Re. This better agreement of the temperature field with the data at the higher Reynolds number is consistent with similar observations for the mean axial velocity and turbulent kinetic energy predictions (Figs. 11 and 14).

*Species mass fraction*

The radial distributions of the major species mass fractions are presented in the Figures 17-26. For flame F3 (Re=24000), mean $CH_4$ concentrations are well captured by the modified TF model (Fig. 17). The PDF model also shows good agreement except for a modest over-prediction at the lower Re. However, RANS based G-eqn predictions show a consistent under-prediction and an apparent discrepancy in the radial mean flame front positions at all the axial locations. This radial shift can be linked with the mean temperature profile (Fig. 15) which also shows radial movement towards centerline. The profile of $CH_4$ curve clearly shows the premixed un-burnt core (plateau region near the centerline), the flame region where methane is consumed (represented by the decay region), and the burnt or outer region where no methane is present. In general, the modified TF model predictions are in excellent agreement with the data, and the level of agreement is even better than that observed with the PDF model. The $O_2$ concentrations are shown in Fig. 19, exhibit distributions similar to the $CH_4$ curve, and the experimental trends are well captured by both the modified TF model the PDF model; however, the G-eqn predictions show similar trends of radial shift in the flame region. The modified TF model appears to over-predict oxygen concentrations in the outer regions. This observation leads to an under-prediction of $H_2O$ in the outer regions as well (Fig. 25). $CO_2$ concentrations with the TF model are also in good agreement with the measurements (Fig. 21) at all X/D locations, and in fact show better agreement compared to other simulations. CO predictions (Fig. 23) show excellent agreement at X/D of 2.5 but under-predict at other X/D locations. This may possibly be due to the lower $O_2$ consumption. However, both the RANS based model simulations, i.e. PDF model and G-eqn model, always significantly over estimate the CO concentrations.

In the case of flame F1 (Re=52000), mean $CH_4$ concentrations are well reproduced by modified TF model (Fig. 18) and show predictions that are similar to the case of flame F3. However, $O_2$ concentrations (Fig. 20) show under predictions at the downstream locations and correlate with the slight over-prediction of $CH_4$. This is happening due to enhanced mixing with cold ambient air which is entrained to a greater extent due to the higher jet velocity, and reduces the $O_2$ consumption. As a consequence, $CO_2$ and $H_2O$ productions also go down as shown in Figs. 22 & 26. Mean CO concentrations are also under predicted in this case (Fig. 24) as at the lower Re. Comparing the results of all the simulations, the modified TF model shows overall good agreement with the data, whereas both the RANS based models (PDF and G-eqn model) consistently overestimate the CO concentrations and underestimate the $CO_2$ production.

**Conclusions**

In the present paper, a modified Thickened Flame model is used to compute the piloted premixed stoichiometric methane-air flame for the Reynolds number Re = 24,000 and 52,000. Detailed results using the modified TF model with thickening factor of 10 (F=10) and pilot temperature of 2005 K have been obtained with a 2-step chemistry model. These results have been compared with the measurements as well as with other RANS based simulation results using a PDF model and a G-equation model.

The modified TF model predictions with 2-step chemistry have been found to be in satisfactory agreement with the experimental data. The mean axial velocity is well predicted for both the flames, i.e. flame F3 (Re=24000) and flame F1 (Re=52000), but shows higher spreading rate at the downstream locations, especially for low Reynolds number case. In the flame region for flame F3 (Re=24000), the mean reaction progress variable appears to be under predicted along the inner flame brush region, but matches well with the data in the case of flame F1 (Re=52000). The turbulent kinetic energy is under-predicted in the vicinity of the centerline in the near field of injection for both the flames, but is generally well captured at the downstream locations. The major species mass fraction predictions are also in good agreement for both the flames, excluding the CO prediction that is consistently under-predicted.

In general, the modified TF model predictions appear to be in reasonable agreement with the data and show improvement in results in the case of high Reynolds number. The RANS based PDF model simulations generally are in reasonable agreement with the data and the TF model except for an over-prediction of the kinetic energy at the lower Re and a significant over-estimation of the CO levels. The other RANS based approach, G-eqn. model predictions, are usually not in good agreement with the data; they show a radial-shift of the flame brush region, an over-estimation of the progress variable and CO at the lower Reynolds number. Moreover, the LES coupled G-eqn predictions also do not show good agreement with the data, especially for the lower Re.

In comparing the computational resources, the modified TF model is less computationally expensive compared to PDF model where the use of a look-up table takes substantial computational input/output (I/O) requirements. However, the modified TF model simulations at this stage are restricted to very few number of chemical kinetics (1 or 2-step chemistry), but the present simulations demonstrate reasonably good agreement with the data.

## Acknowledgements

This work was supported by the Clean Power and Energy Research Consortium (CPERC) of Louisiana through a grant from the Louisiana Board of Regents. Simulations are carried out on the computers provided by LONI network at Louisiana, USA (www.loni.org) and HPC resources at LSU, USA (www.hpc.lsu.edu). This support is gratefully acknowledged.

**References**

Akselvoll, K. and Moin, P. 1996. Large-eddy simulation of turbulent confined co-annular jets. *Journal of Fluid Mechanics*, 315, 387.

Butler, T. D. and O'Rourke, P. J. 1977. A numerical method for two-dimensional unsteady reacting flows. *Proceedings of the Combustion Institute*, 16, 1503.

Charlette, F., Meneveau, C. and Veynante, D. 2002. A Power-Law flame wrinkling model for LES of premixed turbulent combustion, Part I: Non Dynamic formulation and initial tests. *Combustion and Flame*, 131, 159.

Charlette, F., Meneveau, C. and Veynante, D. 2002. A Power-Law flame wrinkling model for LES of premixed turbulent combustion, Part II: Dynamic formulation and initial tests. *Combustion and Flame*, 131, 181.

Chen, Y. C., Peters, N., Schneemann, G. A., Wruck, N., Renz, U. and Mansour, M. S. 1996. The detailed flame structure of highly stretched turbulent premixed methane-air flames. *Combustion and Flame*, 107, 233.

Colin, O., Ducros, F., Veynante, D. and Poinsot, T. 2000. A thickened flame model for large eddy simulation of turbulent premixed combustion. *Physics of Fluids*, 12, 1843.

De, A. and Acharya, S. 2008. Large Eddy Simulation of Premixed Combustion with a Thickened-Flame Approach. *ASME Turbo Expo 2008: Power for Land, Sea and Air*. ASME Paper GT2008-51320.

Duchamp de La Geneste, L. and Pitsch, H. 2000. A level-set approach to large-eddy simulation of premixed turbulent combustion. *Annual Research Briefs*, CTR, Stanford, 105.

Durand, L. and Polifke, W. 2007. Implementation of the thickened flame model for large eddy simulation of turbulent premixed combustion in a commercial solver. *ASME Turbo Expo 2007: Power for Land, Sea and Air*. ASME paper GT2007-28188.

Edwards, J. R. 1997. A low-diffusion flux-splitting scheme for Navier-Stokes calculations, *Computers and Fluids*, 26, 635.

Germano, M., Piomelli, U., Moin, P. and Cabot, W. H. 1991. A dynamic subgrid-scale eddy viscosity model. *Physics of Fluids*, 3, 1760.

Herrmann, M. 2006. Numerical simulation of turbulent Bunsen flames with a level set flamelet model. *Combustion and Flame*, 145, 357.

Ihme, M. and Pitsch, H. 2008. Prediction of extinction and reignition in nonpremixed turbulent flames using flamelet/progress variable model, 1. A priori study and presumed PDF closure. *Combustion and Flame*, 155, 70.

Karypis, G. and Kumar, V. 1999. METIS: A software package for partitioning unstructured graphs, partitioning meshes, and computing full-reducing orderings of sparse matrices, version 4.0. *Technical Report*, University of Minnesota, Department of Computer Science and Engineering.

Kee, R. J., Dixon-Lewis, Warnats, J., Coltrin, M. E. and Miller, J. A. 1986. Technical Report SAND86-8246 (TRANFIT), Sandia National Laboratories, Livermore, CA.

Kee., R. J., Rupley, F. M. and Miller, J. A. 1989. Chemkin-II: A Fortran Chemical Kinetics Package for Analysis of Gas-Phase Chemical Kinetics. Sandia Report, SAND 89-8009B.

Kuo, Kenneth K. 2005. Principles of Combustion, Second edition. John Wiley & Sons. Inc.

Legier, J. P., Poinsot, T. and Veynante, D. 2000. Dynamically thickened flame LES model for premixed and non-premixed turbulent combustion. *Summer Program*, Center for Turbulent Research, Stanford University, 157.

Lindstedt, R. P. and Vaos, E. M. 2006. Transported PDF modeling of high-Reynolds-number premixed turbulent flames. *Combustion and Flame*, 145, 495.

Peters, N. 2000. Tubulent Combustion, Cambridge Univ. Press, London/New York.

Pitsch, H. and Duchamp de La Geneste, L. 2002. Large-eddy simulation of a premixed turbulent combustion using level-set approach. *Proceedings of the Combustion Institute*, 29, 2001.

Poinsot, T. and Veynante, D. 2001. Theoretical and Numerical Combustion, Edwards.

Pope, S. B. 1985. Pdf methods for turbulent reactive flows. *Progress in Energy Combustion Science*, 11, 119.

Prasad, R. O. S. and Gore, J. P. 1999. An evaluation of flame surface density models for turbulent premixed jet flames. *Combustion and Flame*, 116, 1.

Roux, S., Lartigue, G., Poinsot, T., Meier, U., Bérat, C. 2005. Studies of mean and unsteady flow in a swirled combustor using experiments, acoustic analysis, and large eddy simulations. Combustion and Flame, 141, 40.

Selle, L., Lartigue, G., Poinsot, T., Koch, R., Schildmacher, K. U., Krebs, W., Prade, B., Kaufmann, P. and Veynante, D. 2004. Compressible large eddy simulation of turbulent combustion in complex geometry on unstructured meshes. *Combustion and Flame*, 137, 489.

Sagaut, P. 2001. Large Eddy Simulation for Incompressible Flows. Springer-Verlag Berlin, Germany.

Smagorinsky, J. 1963. General circulation experiments with the primitive equations. I: The basic experiment. *Monthly Weather Review*, 91, 99.

Smirnov, A., Shi, S. and Celik, I. 2001. Random Flow generation technique for large eddy simulations and particle-dynamics modeling. *ASME Journal of Fluids Engineering*, 123, 359.

Weiss, J. M. and Smith, W. A. 1995. Preconditioning applied to variable and constant density flows. *AIAA Journal*, 33, 2050.

Williams, F. A. 1985. Combustion Theory, Benjamin/Cummins, Menlo park, CA.

Yeung, P. K., Girimaji, S. S. and Pope, S. B. 1990. Straining and scalar dissipation on material surfaces in turbulence: implications for flamelets. *Combustion and Flame*, 79, 340.

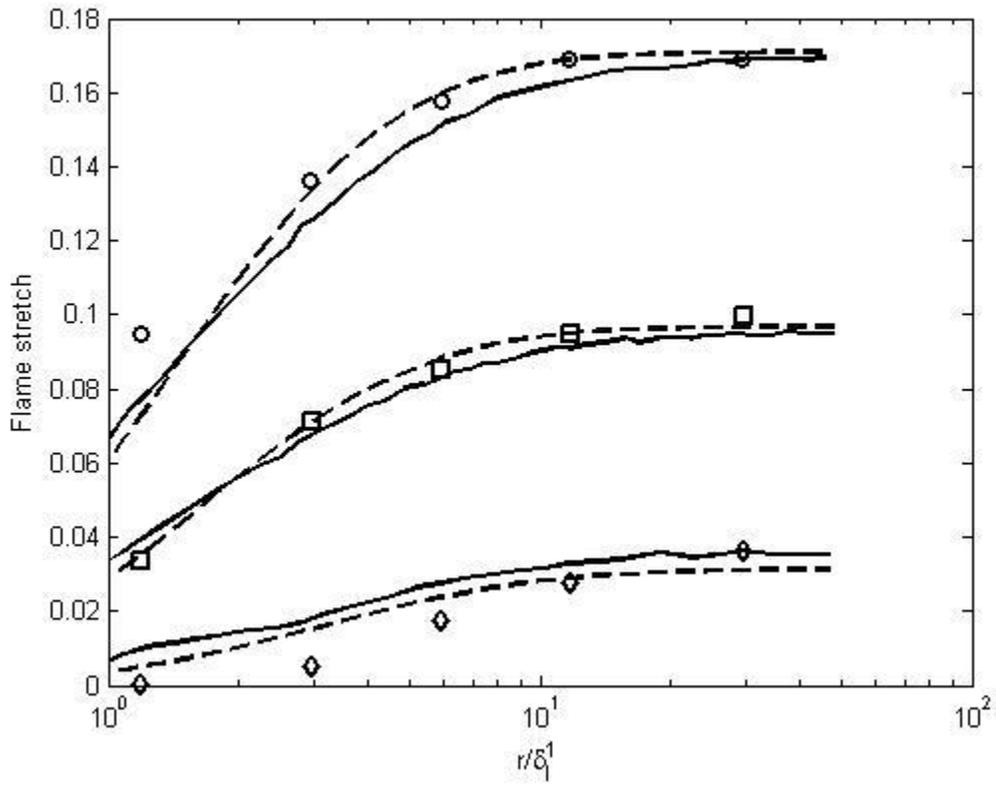

Figure 1. Flame stretch for different length scale ratio. DNS results from Colin et al. (2000): cases An (○), Bn (□), Cn (◊). The $C_n$ function [Eq. 21] (—), The modified $C_n$ function [Eq. 22] (---).

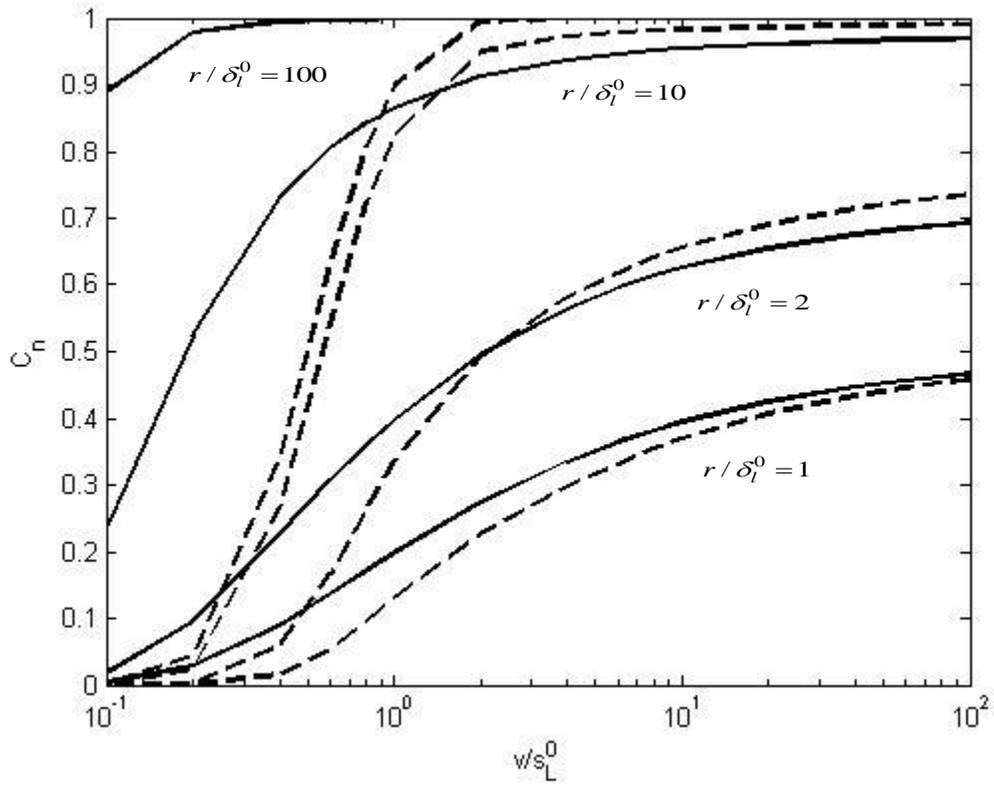

Figure 2. Vortex efficiency functions for different velocity scale ratio. The function [Eq. 21] proposed by Colin et al. (2000) (—), The modified function [Eq. 22] (---).

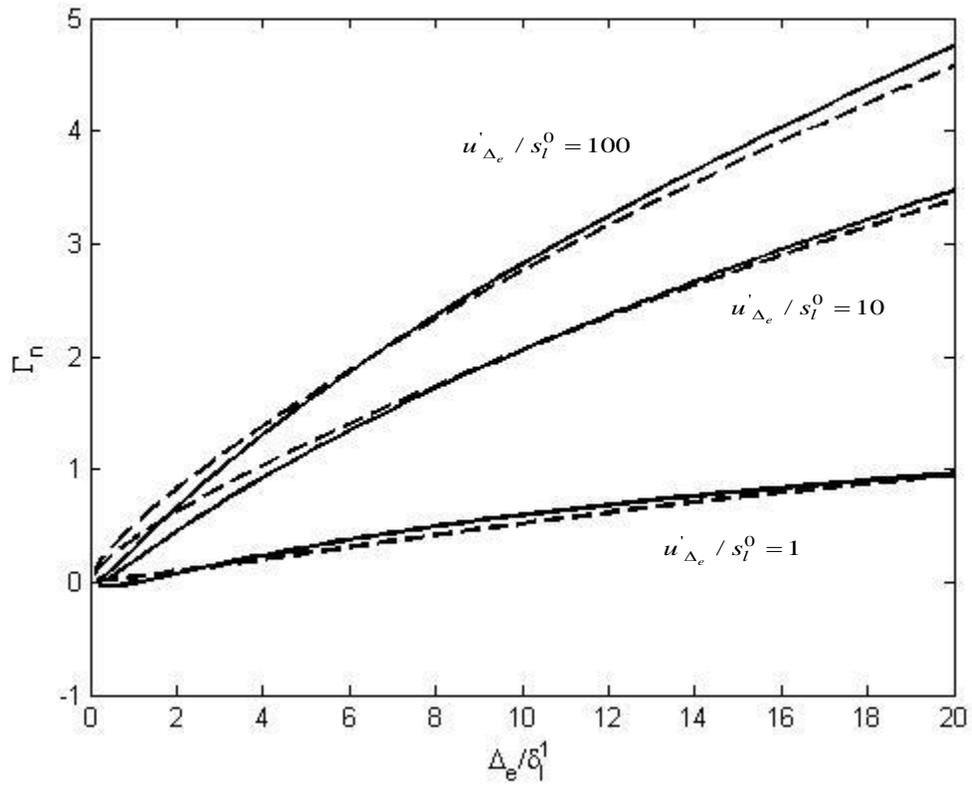

Figure 3. $\Gamma_n$ plotted for $l_t/\delta_L^1 = 100$. Solid lines ( — ) represent the results from the numerical integration of Eq. (26), Dashed lines ( --- ) represent the proposed fit [Eq. 27].

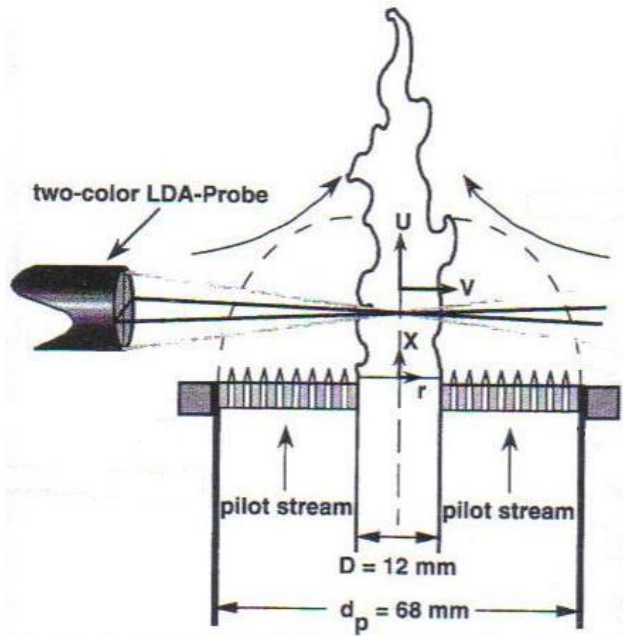

Figure 4. Schematic diagram of the Bunsen burner with enlarge pilot flame (Chen et al., 1996).

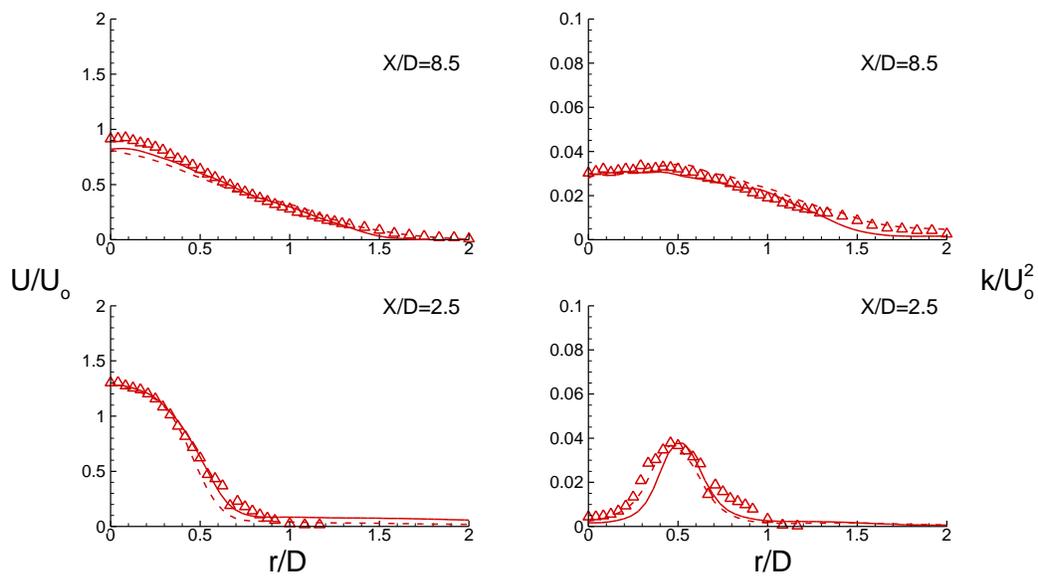

Figure 5. Cold flow (Re=24000): mean axial velocity $U/U_o$ and turbulent kinetic energy $k/U^2_o$, Experimental data is shown by symbols ($\Delta$) and lines are LES results: coarse ( —), fine ( ---).

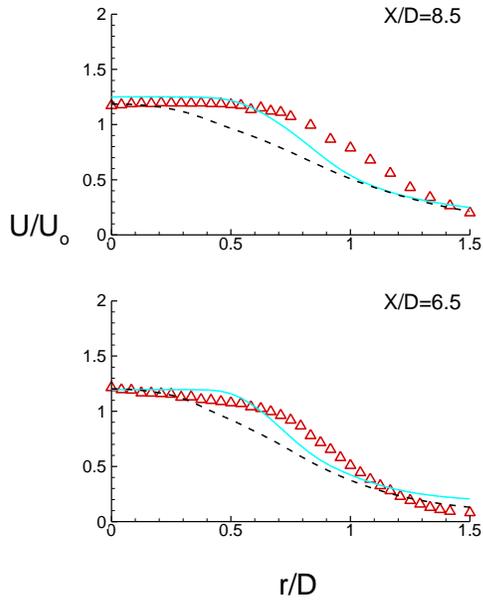
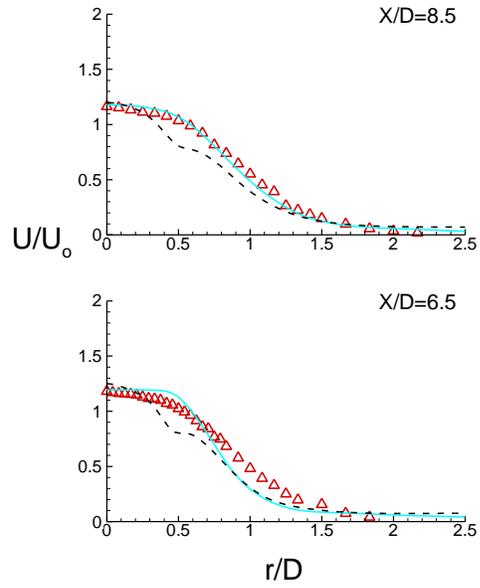

Figure 6. Reacting flow: Mean axial velocity $U/U_o$ (Re=24000). Experimental data ($\triangle$), modified TF model (—), original TF model (---).

Figure 7. Reacting flow: Mean axial velocity $U/U_o$ (Re=52000). Legend-See Fig. 6

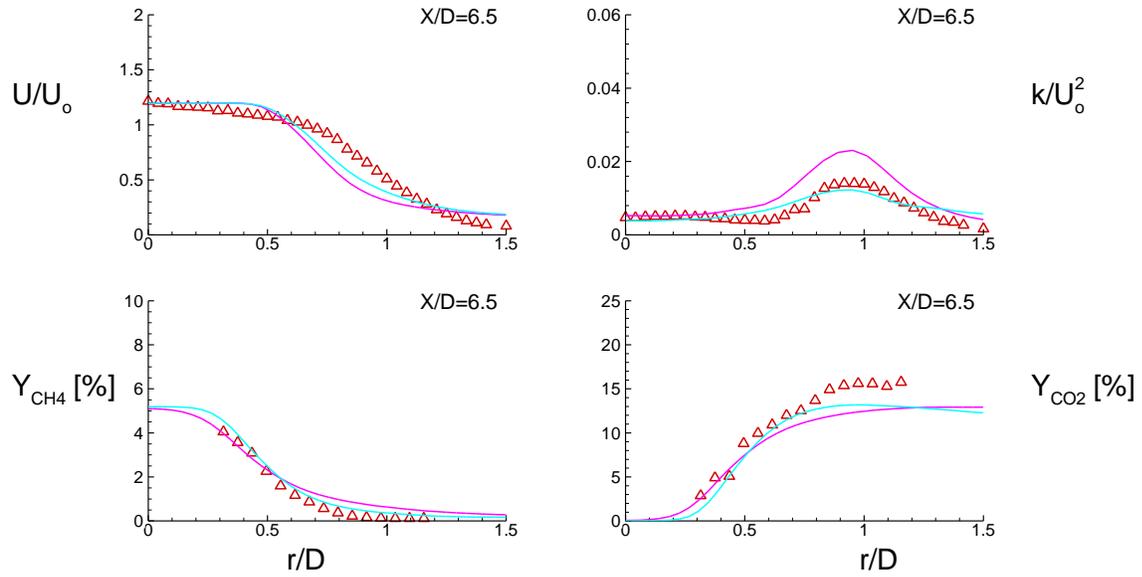

Figure 8. Reacting flow: Mean axial velocity $U/U_o$, turbulent kinetic energy $k/U^2_o$, Mean $CH_4$ concentrations, Mean $CO_2$ concentrations using different thickening factor F (Re=24000). Experimental data (△), F=10 ( — ), F=20 ( — ).

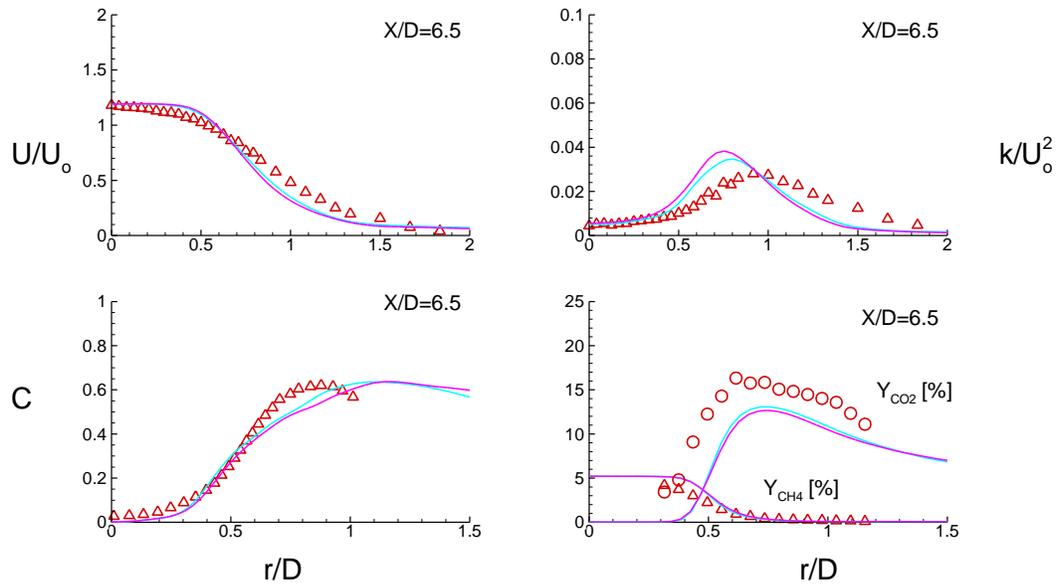

Figure 9. Reacting flow: Mean axial velocity U/U$_o$, turbulent kinetic energy k/U$^2_o$, Mean temperature *C*, Mean CH$_4$ concentrations, Mean CO$_2$ concentrations using different pilot temperature (Re=52000). Experimental data (Δ, ○), 2005K ( — ), 1785K ( — ).

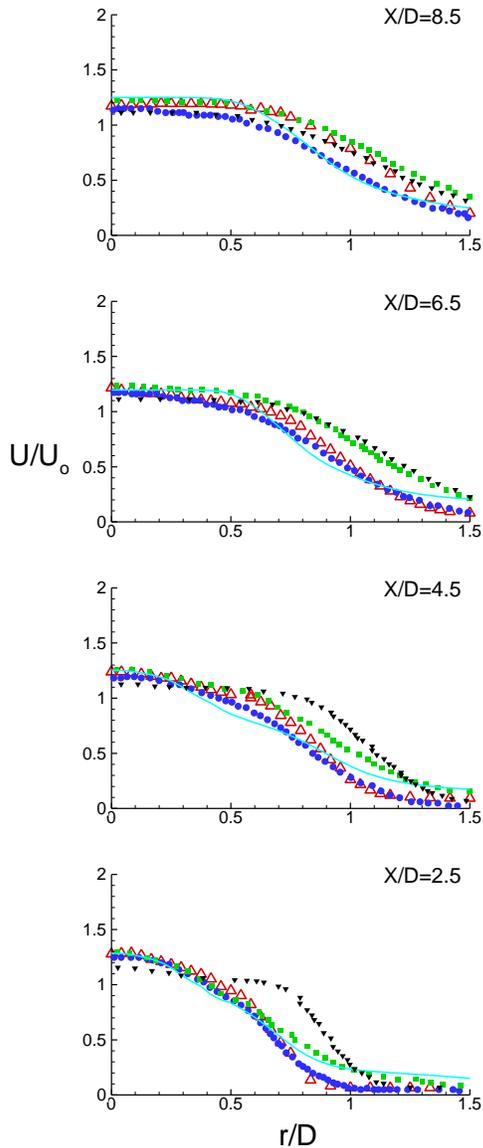
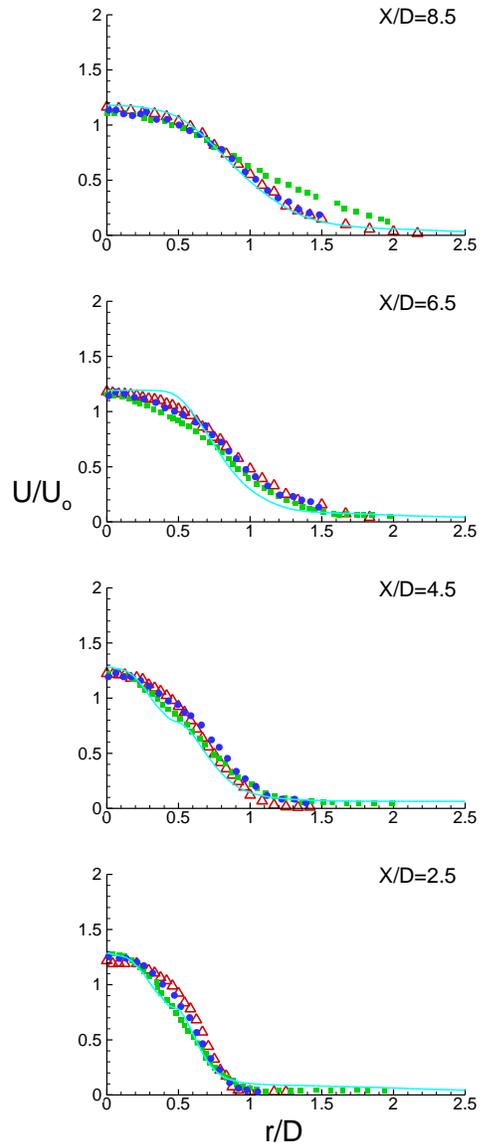

Figure 10. Reacting flow: Mean axial velocity $U/U_o$ (Re=24000). Experimental data ($\triangle$), Lindstedt simulations with RANS based PDF model ($\bullet$), Hermann simulations with RANS based G eqn. model ($\blacksquare$), Duchamp simulations with LES based G eqn. model ($\blacktriangledown$), TF model (—).

Figure 11. Reacting flow: Mean axial velocity $U/U_o$ (Re=52000). Legend-See Fig. 10

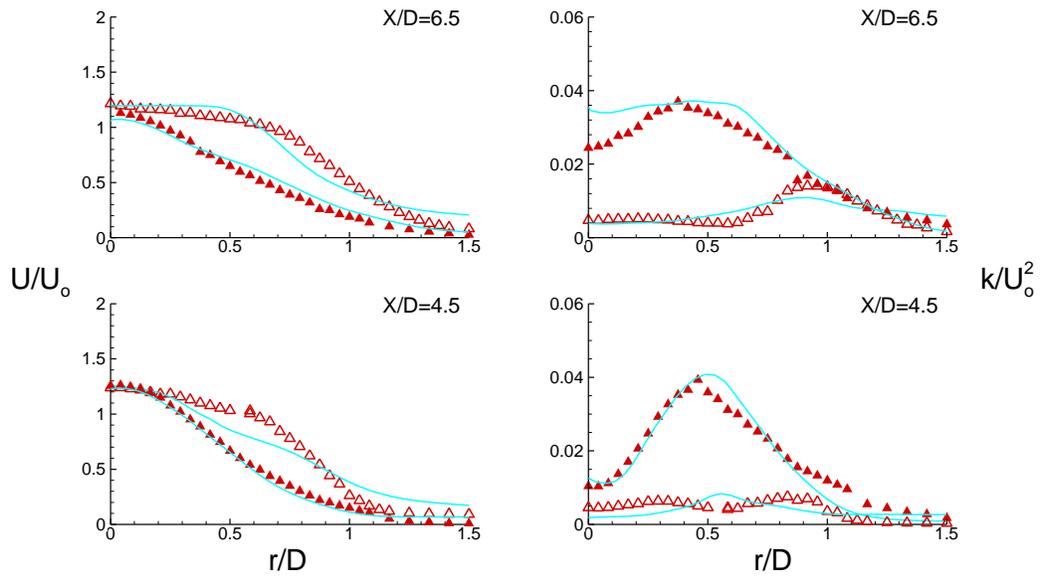

Figure 12. Mean axial velocity $U/U_o$ and turbulent kinetic energy $k/U^2_o$ at Re=24000: Experimental data: reacting flow ($\Delta$), cold flow ($\blacktriangle$); Lines are LES predictions.

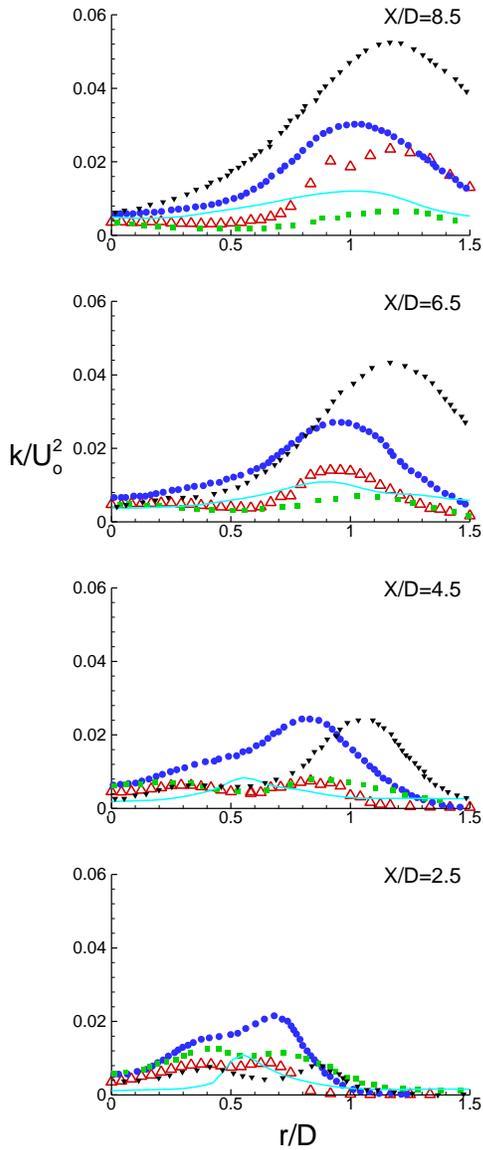
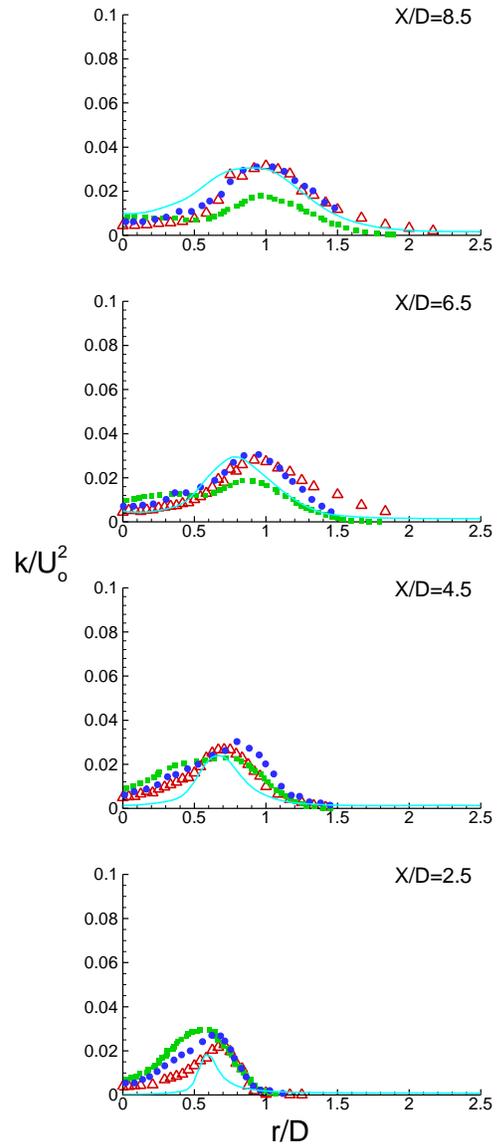

Figure 13. Reacting flow: turbulent kinetic energy $k/U_o^2$ (Re=24000). Legend-See Fig. 10

Figure 14. Reacting flow: turbulent kinetic energy $k/U_o^2$ (Re=52000). Legend-See Fig. 10

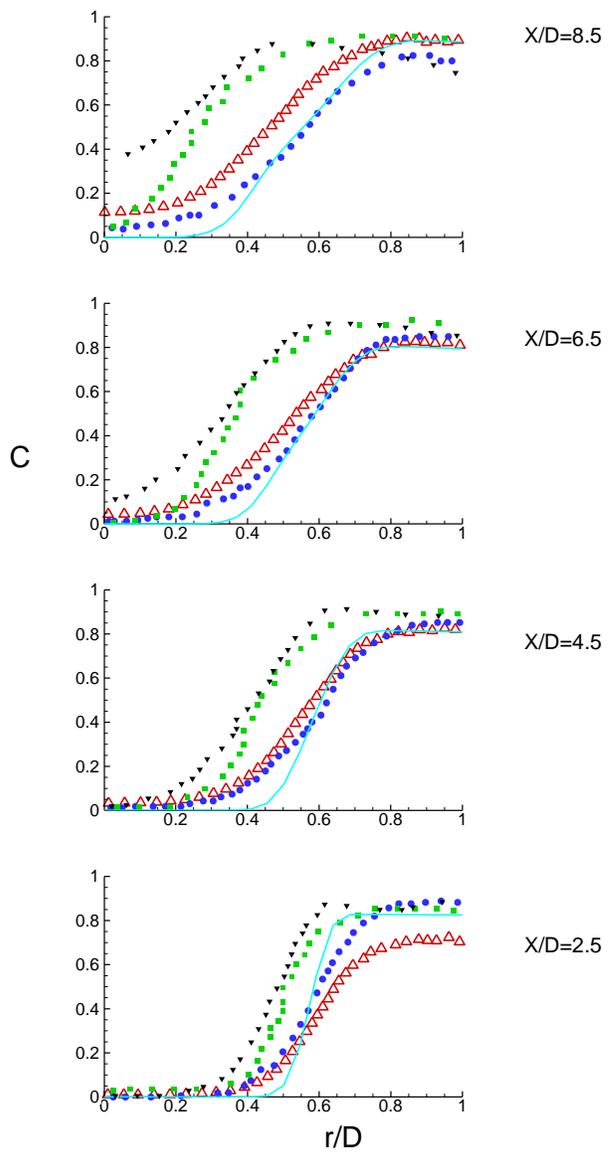

Figure 15. Reacting flow: Mean temperature C (Re=24000). Legend-See Fig. 10

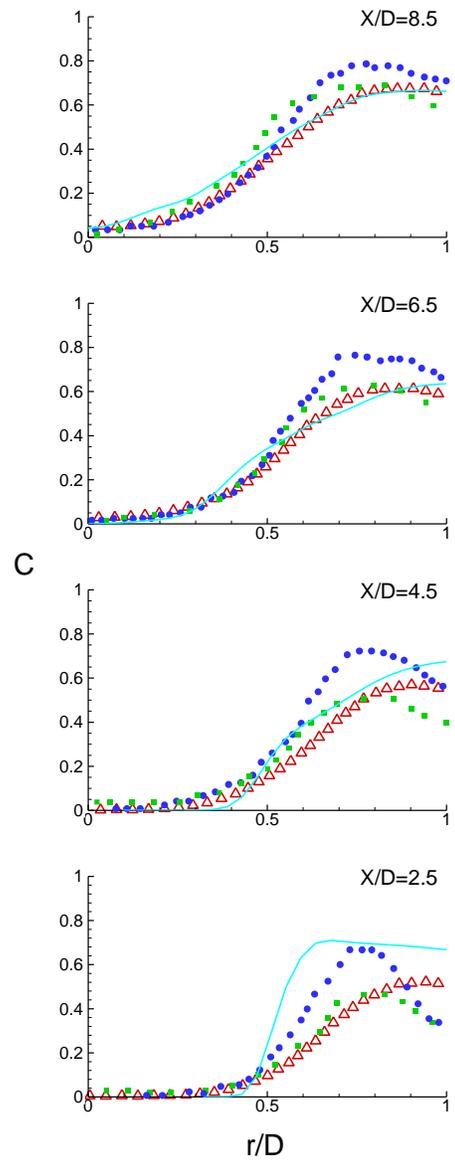

Figure 16. Reacting flow: Mean temperature C (Re=52000). Legend-See Fig. 10

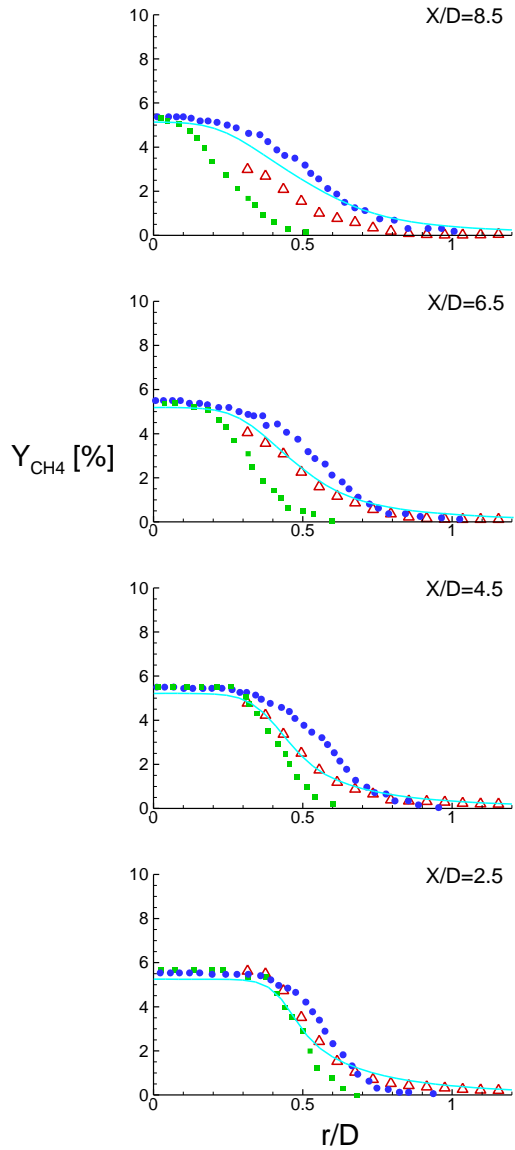
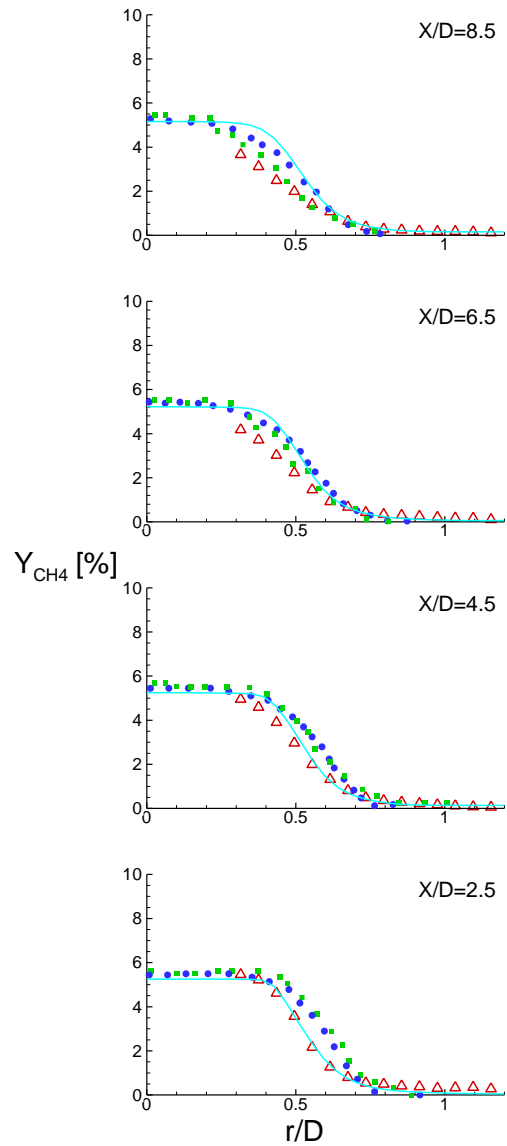

Figure 17. Reacting flow: Mean $CH_4$ concentrations (Re=24000). Legend-See Fig. 10

Figure 18. Reacting flow: Mean $CH_4$ concentrations (Re=52000). Legend-See Fig. 10

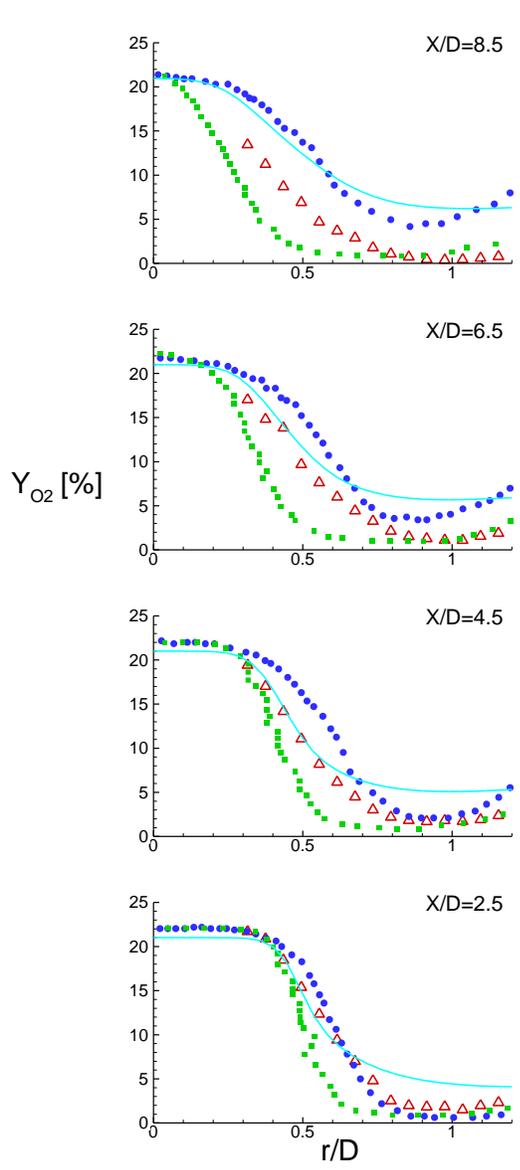
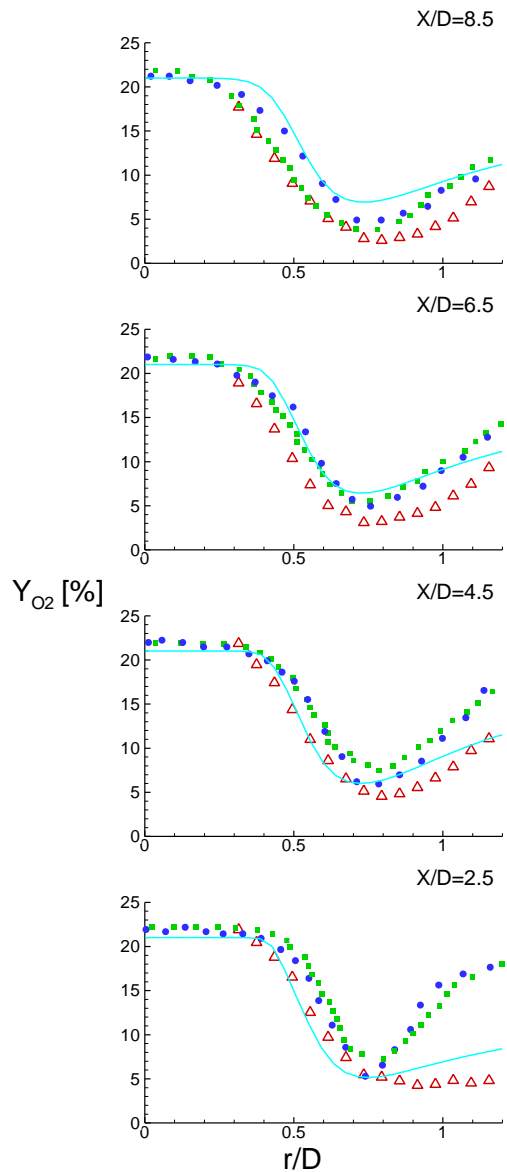

Figure 19. Reacting flow: Mean $O_2$ concentrations (Re=24000). Legend-See Fig. 10

Figure 20. Reacting flow: Mean $O_2$ concentrations (Re=52000). Legend-See Fig. 10

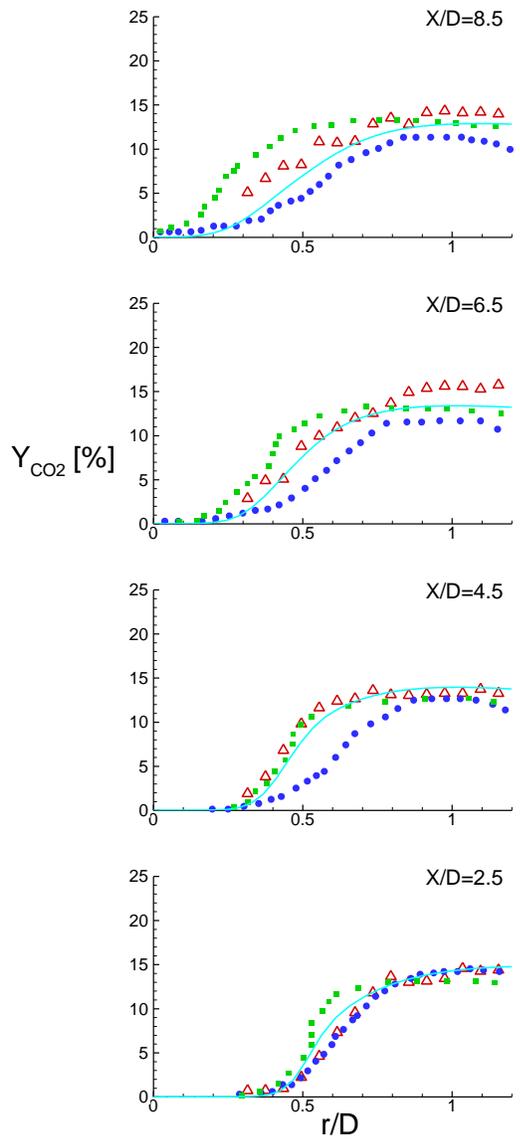
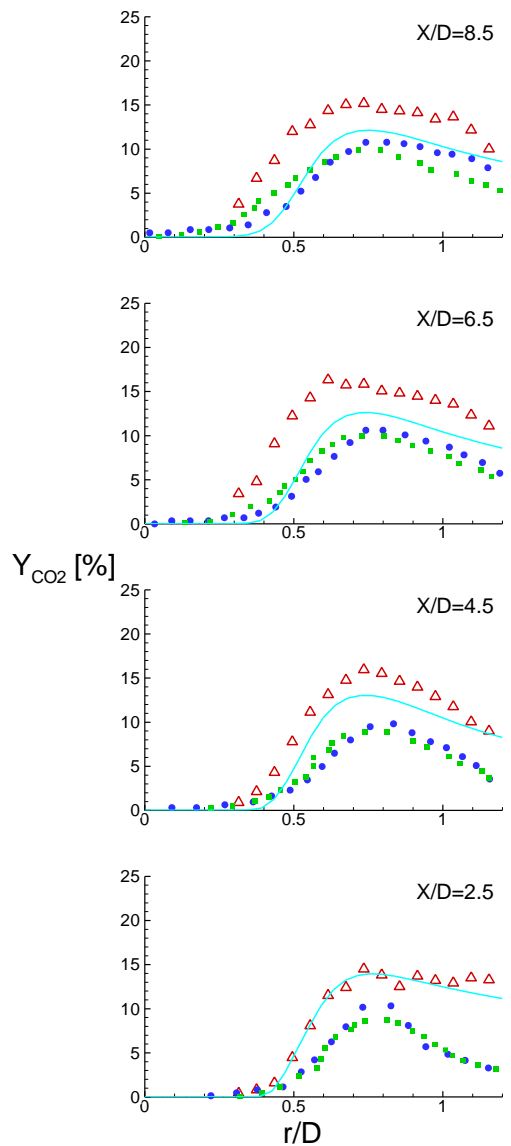

Figure 21. Reacting flow: Mean $CO_2$ concentrations (Re=24000). Legend-See Fig. 10

Figure 22. Reacting flow: Mean $CO_2$ concentrations (Re=52000). Legend-See Fig. 10

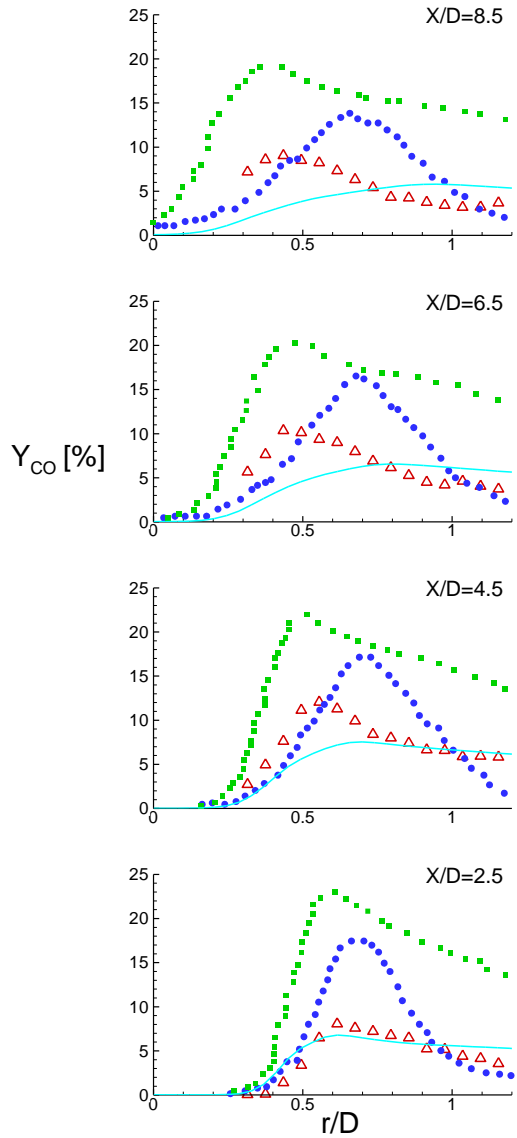

Figure 23. Reacting flow: Mean CO concentrations (Re=24000). Legend-See Fig. 10

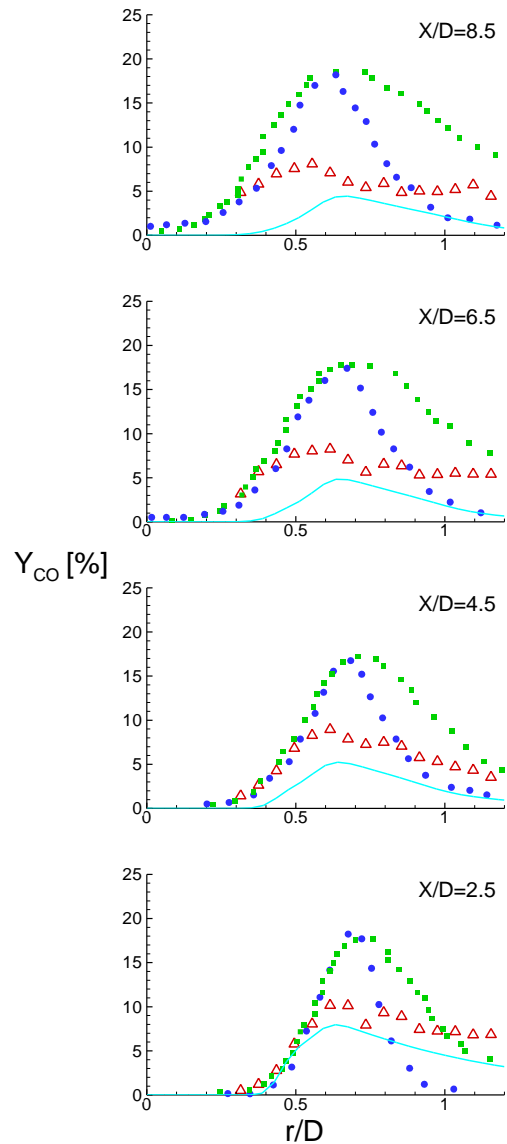

Figure 24. Reacting flow: Mean CO concentrations (Re=52000). Legend-See Fig. 10

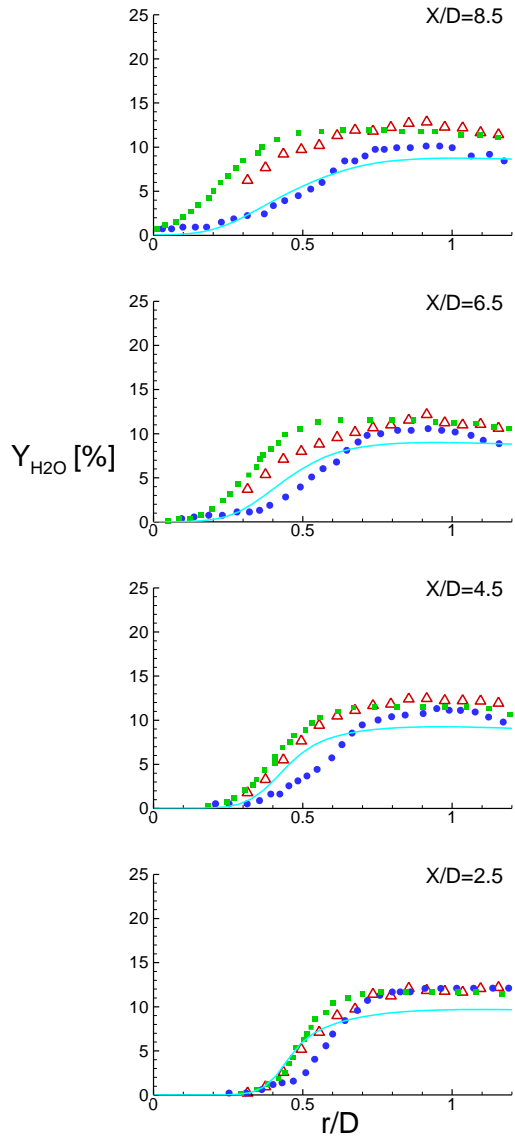
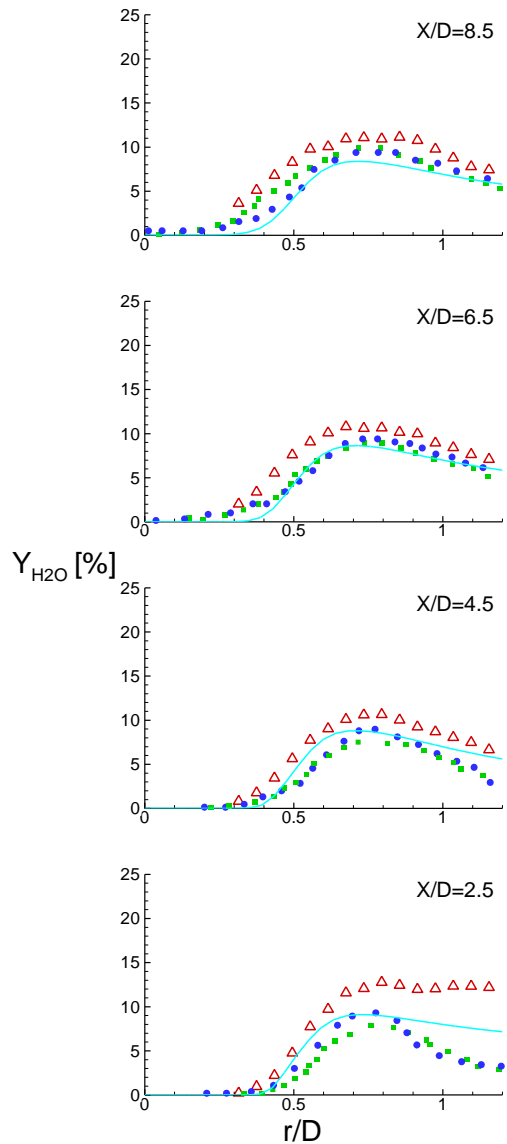

Figure 25. Reacting flow: Mean $H_2O$ concentrations (Re=24000). Legend-See Fig. 10

Figure 26. Reacting flow: Mean $H_2O$ concentrations (Re=52000). Legend-See Fig. 10